\documentclass[12pt]{iopart}
\usepackage{float}
\usepackage{iopams}
\usepackage{color,soul}

\usepackage[normalem]{ulem} 
\usepackage{xcolor}         
\usepackage{tikz}

\newcommand{\Var}{\mathop{\mathrm{Var}}}

\definecolor{mahogany}{RGB}{153, 50, 25}

\newcommand{\floor}[1]{\mathop{\lfloor{#1}\rfloor}}

\begin{document}
\title[New probability distribution describing emergence in state space]
{New probability distribution describing emergence in state space}

\author{Roozbeh H. Pazuki$^1$}
\eads{rpazuki@gmail.com}

\author{Henrik Jeldtoft Jensen$^{1,2}$}
\eads{\mailto{h.jensen@imperial.ac.uk}}

\vspace{.5cm}

\address{$^1$Centre for Complexity Science and Department of Mathematics, Imperial College London, South Kensington Campus, SW7 2AZ, UK}
\address{$^2$Institute of Innovative Research, Tokyo Institute of Technology, 4259, Nagatsuta-cho, Yokohama 226-8502, Japan}

\begin{abstract}
We revisit the pairing model of state spaces with new emergent states introduced in \jpa 51 375002, 2018.

We facilitate our analysis by introducing a simplified pairing model consisting of balls able to form pairs but without any internal structure.

For both the simplified and the original model we compute exactly the probability distribution for observing a state with $n_p$ pairs. We show this distribution satisfies a large deviation principle with speed $n \ln(n)$. We present closed form expressions for a variety of statistical quantities including moments and marginal distributions.
\end{abstract}
\\

\section{Introduction}
\label{Introduction}
The focal point of complexity science is the study of the emergent structures generated by the interactions between the components of  a given system. Depending on the nature of these interactions, the  number of available states $W(N)$ accessible to the fully interacting system will exhibit different functional dependencies on the number of components $N$. 

In statistical mechanics the state space of the $N$ component system is given by the Cartesian product of the state space of the individual components. For this reason $W(N)$ will be exponential in $N$, or asymptotically it grows as $W(N) \sim O(k^N)$ for a real, positive constant $k$, such as in the Ising model where $W(N)=2^N$. If the interdependence between the components is able to freeze out some of the Cartesian product states and makes the states inaccessible, $W(N)$ may grow slower than   exponentially; some examples of this case are described in \cite{hanel2011}.

In contrast, if new collective states become possible as a result of the inter-component  interaction, $W(N)$ will grow faster than exponentially. One may, e.g., think of the formation of hydrogen molecules out of hydrogen atoms. Hydrogen molecules exhibit properties that are not intrinsic to hydrogen atoms. Taking into account the emergent properties of hydrogen molecules as new collective states, the state space of a compound system -- atoms and molecules -- will grow faster than an exponential function.

One all inclusive definition of emergence does not exist. Here we focus on situations where interaction generates entirely new states different from the Cartesian combination of single particle states. From this perspective the hydrogen molecule is a new emergent state. To do this we study the statistics of a minimal model of such cases, namely the pairing model introduced in \cite{jensen2018}. The pairing model has been studied in the context of relating the $N$ dependence of $W(N)$ to generalised entropies in a number of publications, see \cite{jeldtoft2018, korbel2018,tsallis2019,korbel2019,vigneaux2019,ilic2019,Tempesta2020b,korbel2020,balogh2020}. One may think of the components of the pairing model as coins. Each coin can either show head or tail. In addition to the single component states, head and tail, two coins can form a paired state. 

As a result $W(N)$ grows faster than exponentially and the dependence on $N$ is controlled by a recursive equation, which we use to compute statistics of the paired and unpaired components. This allows us to present large deviation expressions with a speed different from $N$, namely of the form $P_N(x) \asymp \exp[-N \ln N I(x)]$. 

From now on, to make our formulae more readable we will use lower case $n$ to denote the system size to reduce clutter. We will introduce two probability distributions, $P_{n}(n_p)$ and $P_{n}(n_p, n_h)$, where the number of coins showing head is denoted by $n_h$ and the number of pairs by $n_p$. 
We point out that most of the properties of these distributions can be computed in  closed form thereby facilitating statistical modelling and further analytical investigation. 

Moreover, we will show that both distributions $P_{n}(n_p)$ and $P_{n}(n_p, n_h)$  satisfy a large deviation property. We recall that a probability distribution satisfies a Large Deviation Property (LDP) \cite{Ellis85} 
if the limit
\begin{equation}
\lim_{n \rightarrow \infty} -\frac{1}{a_{n}} \ln P_{n}(X_{n})
\end{equation} 
exists for a random variable $X_n$, a sequence of distributions $P_n$, and a sequence of positive numbers $a_n$, called speed, for $ n \in \{1,2, \dots\}$ that tends to $\infty$.

The large deviation speeds that we encounter in standard statistical mechanics corresponding to state spaces sizes $W(n)$  exponential  in  $n$, are usually linear $a_n = n$. However, non-linear speed is studied  in the large deviation literature. See for example \cite{LOWE1996, chatterjee2011, lubetzky2015}.

The $n\ln n$ speed becomes particularly interesting when one realises that a system of pairing coins corresponds to a graph where each node has either degree zero or one. This is the simplest class of the more general class of non-trivial sparse networks, for which the form of the  LDP still remains open though considerable progress has been achieved in recent years \cite{chatterjee2012, chatterjee2016, cook2020} (\textit{e.g.} see \cite{chatterjee2016} for an introduction).

In the sparse graphs case, quadratic speed, or $a_n = n^2$,  is needed to ensure the existence of non-trivial large deviation bounds  \cite{chatterjee2016}. However, these bounds do not fully solve ultra-sparse networks problem. For instance, for a Erd{\"o}s-R{\'e}nyi graph $G(n, p)$ in which $p \rightarrow 0$ and $n \rightarrow \infty$ and sampled random networks are ultra-sparse, the problem has not been fully resolved.  Because of the indicated resemblance of the pairing model and ultra-sparse networks, the results of the present paper are relevant to the LDP for sparse graphs, and in particular, suggest that the speed $n \ln n$ may be of importance to such networks. 

To find the large deviation distribution of $P_{n}(n_p)$ and $P_{n}(n_p, n_h)$ for a random configuration of size $n$, let us denote the ratio of the number of pairs by $m_n$ as
\begin{equation}
m_{n} \equiv \frac{2n_p}{n} = \frac{1}{n} \sum_{i=0}^{n} \delta_{X_i, p}, \qquad 0 \le m_{n} \le 1,
\end{equation}
where $\delta_{X_i, p}$ is the Kronecker delta and equal to one if $X_i$ is in a pair state, and the ratio of the number of head, namely $s_n$, states as
\begin{equation}
s_{n} \equiv \frac{n_h}{n-2n_p} = \frac{1}{n-2n_p}  \sum_{i=0}^{n} \delta_{X_i, 1}, \qquad 0 \le s_{n} \le 1,
\end{equation}
whereas $\delta_{X_i, 1}$ is equal to one for head states. For $m_n$ and $s_{n}$, we shall derive the large deviation probabilities $P_n(m_{n})$ and $P_n(m_{n}, s_{n})$ with speed $n \ln n$ in the forthcoming sections.

The statistics also contain quantities described by a LDP with the usual linear speed $n$. This happens when the relative abundance of non-paired to paired elements is keep fixed while $n$ increases. 

The rest of the paper is organised as follows. The next section describes the pairing model and presents the recursive relation for the state space volume $W(n)$ and discusses a number of random variables and their probability distributions. Sec. \ref{Bino-like} defines a family of binomial-like distributions to formalise our discussion of the large deviation expressions presented in Sec. \ref{Large-devi}. Sec. \ref{Limi-case} presents results for the case where the system size grows in relation to number of interactions. Sec. \ref{Closed-form} derives moments of the distributions in closed form. This is followed by a discussion of expressions for marginal distributions in Sec. \ref{Marginal-dist}. We present the summary in the last section.

Finally, for completeness we include a number of tables at \ref{appendix:tables} to list the distributions and their properties. See tables  \ref{table:p_2n_n_p}, \ref{table:p_2n_n_p_n_h}, \ref{table:p_2n_epsilon} and \ref{table:p_2n_epsilon_2}.

\section{The model and its recursive state space volume}
\label{Model}
Think of the components of the pairing model as coins. Each coin can either show head or tail or form a paired state with another coin. The state space volume of $n$ coins is determined by the following recursive rule \cite{jensen2018}
\begin{equation}
   W(n) = 2 W(n-1) + (n-1)W(n-2).
   \label{eq:recursive}
\end{equation}
The asymptotically leading behaviour of the solution to this equation is given by 
\cite{jensen2018}
\begin{equation}
  W(n) \sim \frac{1}{\sqrt{2}\rme} \left(\frac{n}{\rme}\right)^{n/2} \rme^{2\sqrt{n}}.
\end{equation}

Despite the model's simplicity, to the best of our knowledge, the probability distributions describing the  configurations in terms of their number of heads, tails and paired coins have not been reported in the literature. 

It turns out to be instructive to introduce a variation of the model, which is even simpler than the original pairing model. Let us refer to the original version introduced in \cite{jensen2018} as\\

\noindent\underline{Coins Model ($C$-Model)}: consisting of $n$ coins. Each coin can be in one of two states, head or tail, or in a paired state together with another coin. The $C$-Model is introduced and discussed in \cite{jensen2018}.\\

\noindent\underline{Balls Model ($B$-Model)}: consisting of $n$ structureless objects, we will call them balls. Each ball can occupy one single particle state only -- stand-alone state--, but any two balls are able to combine and form a paired state. The $B$-Model is introduced for the first time in this paper.

It is illuminating to analyse the pairing in a way similar to the usual binomial analysis of Bernoulli variables. Where the binomial distribution corresponds to binary random variables, we will also look at \textit{binomial-like} distributions suitable for the analysis of random variables that besides the binary single component state are able to form pairs. In the next section we do this analysis starting with the simpler $B$-model before we turn tot he $A$-model.

\section{A Binomial-like distribution}
\label{Bino-like}

\subsection{Model $B$-Model}
Let $\mathcal{S}_n$ denotes the set of all possible configurations for the $B$-Model. The state space volume of $n$ balls is determined by the following rule
\begin{equation}
\label{eq:recursive.2}
W(n) = W(n-1) + (n-1)W(n-2).
\end{equation} 

The number of pairs denoted by $n_p$ is a random variable, the ensemble of configurations of the $B$-Model, $\mathcal{S}_n$. From the outset, we assume equal probability among configurations that consist of the same number of pairs.
A specific configuration with $i$ pairs by $c_i$, $i \in \{0, 1, \dots, \lfloor \frac{n}{2}\rfloor\}$.

Partitioning $\mathcal{S}_n$ by the number of pairs introduces the following disjoint subsets
\begin{equation}
S_i := \{ c_i \in \mathcal{S}_n:  c_i \mbox{ has } i \mbox{ pairs} \},
\end{equation}
such that
\begin{equation}
\mathcal{S}_n = \bigcup_{i=0}^{\lfloor \frac{n}{2}\rfloor} S_i	, \qquad S_i \bigcap S_j = \emptyset, \qquad i \ne j. 
\end{equation}
In other words, $S_i$ is a subset of configurations with $i$ pairs.

Next, let $p_i$ denote the probability of the event set $S_i$
\begin{equation}
 p_i\equiv  P_n(S_i) = P_n(n_{p} = i)  \qquad 0 \le p_i \le 1.
\end{equation}
Moreover, since $p_i$ are probabilities the normalization condition requires
\begin{equation}
\label{eq:normalization.1}
\sum_{i=0}^{\lfloor \frac{n}{2}\rfloor} p_i = 1.
\end{equation}
The cardinality of the subset $S_i$ is 
\begin{equation}
\left| S_i \right| = {n \choose 2i} (2i-1)!!.
\end{equation}
Note that the double factorial counts the number of distinguishable pairs. In the case of uniform distribution, we must have 
\begin{equation}
\label{eq:p_i_as_ratio}
   p_i = P_n(S_i) = \frac{|S_i|}{|\mathcal{S}_n|} = \frac{{n \choose 2i} (2i-1)!!}{W(n)},
\end{equation}
In contrast, the probability for observing a specific configuration $c_i \in S_i$ defines by 
\begin{equation}
\label{eq:q_i_equive_to_P_c_c_i}
q_i\equiv P_n(c_i).
\end{equation}
Using the assumption of equal probability among configurations with the same number of pairs, $p_i$ must be equal to $q_i$ times the cardinality of the subset $S_i$. Thus
\begin{equation}
P_n(S_i) = |S_i| \; P_n(c_i) \Rightarrow p_i = {n \choose 2i} (2i-1)!! \; q_i.
\end{equation}
Consequently, the normalization condition (\ref{eq:normalization.1}) reads as
\begin{equation}
\label{eq:normalization_q_i}
\sum_{i=0}^{\lfloor \frac{n}{2}\rfloor} {n \choose 2i} (2i-1)!! \; q_i = 1.
\end{equation}

\begin{figure}[H]
	\centering
	\includegraphics[width=.8\textwidth]{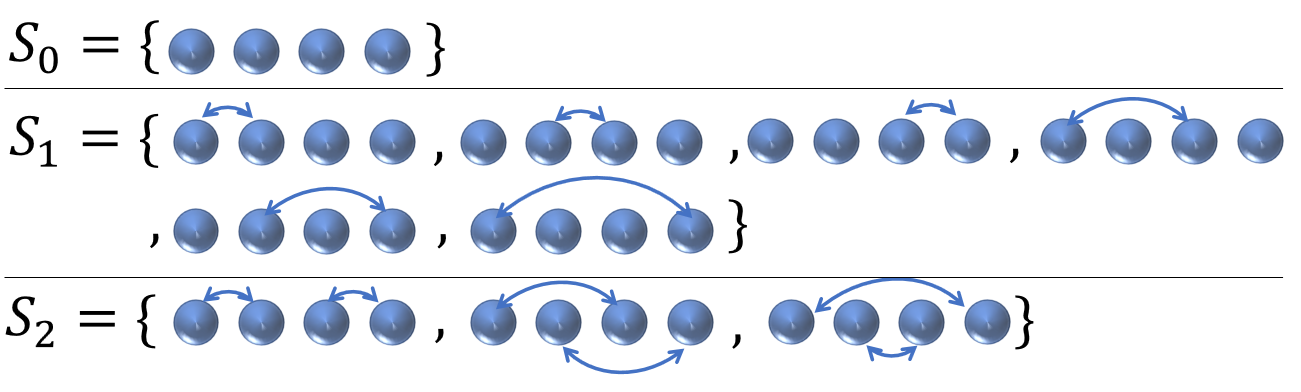} 	
	\caption{Partitioning $\mathcal{S}_4$ to three disjoint subsets. The arrows show the pairing between balls.} 
	\label{fig:balls_s4}	
\end{figure}

Fig. \ref{fig:balls_s4} represents $\mathcal{S}_4$ partitions. Tables \ref{table:s_4_uniform} and \ref{table:s_4_example} show two examples of possible probability distributions.

\begin{table}[!htb]
	\begin{minipage}{.5\linewidth}
		\begin{table}[H]
			\caption{\label{table:s_4_uniform} $\mathcal{S}_4$ Uniform distribution.}
			\begin{indented}
				\item[]\begin{tabular}{@{}lll}
					\br
					$p_i$& $\qquad\qquad$ & $q_i$ \\
					\br
					$p_0 = \frac{1}{10}$& & $q_0 = \frac{1}{10}$\\  			
					$p_1 = \frac{6}{10}$& & $q_1 = \frac{1}{10}$\\
					$p_2 = \frac{3}{10}$& & $q_2 = \frac{1}{10}$\\
					\br
				\end{tabular}
			\end{indented}
		\end{table}
	\end{minipage}%
	\begin{minipage}{.5\linewidth}
		\begin{table}[H]
			\caption{\label{table:s_4_example} Another $\mathcal{S}_4$ distribution.}
			\begin{indented}
				\item[]\begin{tabular}{@{}lll}
					\br
					$p_i$& $\qquad\qquad$ & $q_i$ \\
					\br
					$p_0 = \frac{1}{3}$& & $q_0 = \frac{1}{3}$\\  			
					$p_1 = \frac{1}{3}$& & $q_1 = \frac{1}{3 \times 6}$\\
					$p_2 = \frac{1}{3}$& & $q_2 = \frac{1}{3 \times 3}$\\
					\br
				\end{tabular}
			\end{indented}
		\end{table}
	\end{minipage} 
\end{table}

\subsection{Model $C$-Model}
The set $\mathcal{S'}_n$ denotes the set of all possible configurations for the $C$-Model, and for $n$ coins the cardinality of $\mathcal{S'}_n$ is introduced recursively in (\ref{eq:recursive}).
Recall that for the $B$-Model, we assumed equal probability for configurations in $S_i$. Similarly, for the $C$-Model, the number of pairs in each configuration partitions $\mathcal{S'}_n$ to disjoint subsets. In addition, each configuration has $(n-2i)$ coins that are in non-paired state, and $n_h$ is a random variable that denotes the number of coins in the head state
\begin{equation}
n_{h} \in \{0, 1, \dots, (n-2i)\}.
\end{equation}

Let us denote by $j$, the numbers of head states in a specific configuration $c_i$. For the $C$-Model, we assume equal probability among configuration with the same number of pairs and head states.
Consequently, to obtain the probability distribution for the $C$-Model, when for a single coin the probability of observing a head is $\rho$ and a tail state is $1-\rho$, this new parameter must be included either.

Furthermore, we denote by $S_{ij}$ disjoint subsets of $S_i$ with $j$ coins in head state ($n_{u} = j$), such that
\begin{equation}
S_{ij} := \{ c \in S_i:  c \mbox{ has } j \mbox{ coins in head state} \},
\end{equation}
and
\begin{equation}
S_i = \bigcup_{j=0}^{n-2i} S_{ij}	, \qquad S_{ij} \bigcap S_{ik} = \emptyset, \qquad j \ne k. 
\end{equation}

Given $S_{ij}$ is a subset of $S_i$, we define the conditional probability of event $S_{ij}| S_i$, or probability of observing configurations with $j$ heads, given $i$ pairs
\begin{equation}
P_n(S_{ij}  | S_i) \equiv P_n(n_h = j | n_p = i) = p_{j|i}, \qquad \forall i: \quad 0 \le p_{j|i} \le 1.
\end{equation}
There are ${n-2i \choose j}$ distinct configurations for selecting $j$ heads among $n-2i$ non-paired coins, and therefore $p_{j|i}$ is 
\begin{equation}
p_{j|i} = {n-2i \choose j} \rho^{j} (1-\rho)^{n-2i-j}, \qquad 0 \le \rho \le 1.
\end{equation}
Note that the normalization condition for the conditional distribution, given $i$, satisfies
\begin{equation}
  \sum_{j=0}^{n-2i} p_{j|i} = (\rho + 1 - \rho)^{n-2i} =1.
\end{equation}

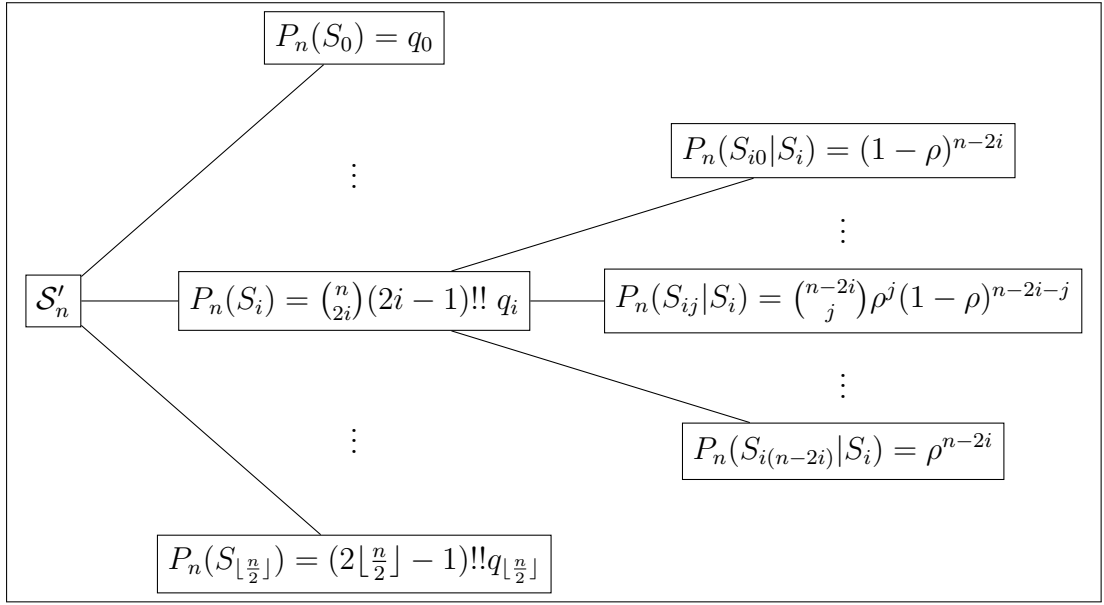
\begin{figure}[H]
	\framebox{
     \begin{tikzpicture}
		\tikzstyle{level 1}=[level distance=40mm,sibling distance=35mm]
		\tikzstyle{level 2}=[level distance=65mm,sibling distance=20mm]
		\node[rectangle,draw] {$\mathcal{S}'_n$}  
		[grow=right]
		child {node [grow=right, rectangle,draw] (a) {$P_n(S_{\lfloor \frac{n}{2}\rfloor})=(2\lfloor \frac{n}{2}\rfloor-1)!! q_{\lfloor \frac{n}{2}\rfloor}$}	   	
		}
		child {node [grow=right, rectangle,draw] (b) {$	P_n(S_{i}) = {n \choose 2i} (2i-1)!! \; q_i$}
			child {node [rectangle,draw] (ba) {$P_n(S_{i(n-2i)}| S_{i})=\rho^{n-2i}$}	   	
			}
			child {node [rectangle,draw] (bb) {$P_n(S_{ij}| S_{i})={n-2i \choose j} \rho^{j} (1-\rho)^{n-2i-j}$}	   	
			}
			child {node [rectangle,draw] (bc) {$P_n(S_{i0}| S_{i})=(1-\rho)^{n-2i}$}	   	
			} 		   	
		}
		child {node [grow=right, rectangle,draw] (c) {$P_n(S_{0})= q_0$}	   	
		}
		;
		\path (a) -- (b) node [midway] {\vdots};
		\path (b) -- (c) node [midway] {\vdots};
		\path (ba) -- (bb) node [midway] {\vdots};
		\path (bb) -- (bc) node [midway] {\vdots};
		\end{tikzpicture}
	}
	\caption{The pairing coins probability tree.}
	\label{fig:prob.tree}
\end{figure}

Finally, the probability of event $S_{ij}$, or observing configurations with $i$ pairs and $j$ heads, is
\begin{equation}
P_n( S_{ij}) \equiv P_n(n_h = j , n_p = i) = p_{ij},
\end{equation}
and according to the probability chain rule it can be written as
\begin{equation}
p_{ij} = p_{i} \times p_{j|i} = {n \choose 2i} (2i-1)!! \; q_i \left[{n-2i \choose j} \rho^{j} (1-\rho)^{n-2i-j}\right].
\end{equation}
Hence, the probability of observing a specific configuration $c_{ij} \in S_{ij}$ with $i$ pairs and $j$ heads must be
\begin{equation}
P_n( c_{ij} ) = q_i \rho^{j} (1-\rho)^{n-2i-j},
\end{equation}
since the cardinality of $S_{ij}$ is ${n \choose 2i} (2i-1)!! \; {n-2i \choose j}$.
This argument is schematically represented in figure (\ref{fig:prob.tree}) as a probability tree. 

\subsection{Closed form of $p_{i}$ and $p_{ij}$}
We shall see, aside from $\rho$, or head state parameter, one parameter is enough to describe the pairing probabilities, $q_i$. 

\begin{figure}[h]
  \begin{tikzpicture}
	\begin{scope}[yshift=-180,yslant=0.5,xslant=-1]
	\filldraw[black!10,very thick] (0.5,1) rectangle (5,7);
	\end{scope}

	\begin{scope}[rotate around = {-5:(0,0,0)}]

	\shadedraw [ball color=gray] (3.55,2.5,12.5) circle(0.25);
	\shadedraw [ball color=gray] (3.59,2.01,12.5) circle(0.25);
	\draw[-latex,thick](-2,-1.5)node[above]
	{$\mathsf{paired}: q_1$} to[out=-90,in=180] (-1.5,-2.5);

	\shadedraw [ball color=white] (2,1,12.5) circle(0.25);
	\draw[-latex,thick](-0.5,-4.5)node[above]
	{$\mathsf{stand}$-$\mathsf{alone}: q_0$} to[out=-90,in=-90] (-2.2,-4.3);

	\shadedraw [ball color=white] (2.7,.7,12.5) circle(0.25);
	\draw[-latex,thick](-4,-3)node[above]
	{$\mathsf{stand}$-$\mathsf{alone}: q_0$} to[out=-90,in=180] (-3,-4);

	\end{scope}	
	\end{tikzpicture}
    \caption{Two balls in paired or stand-alone state.}
	\label{fig:two.coins}
\end{figure}
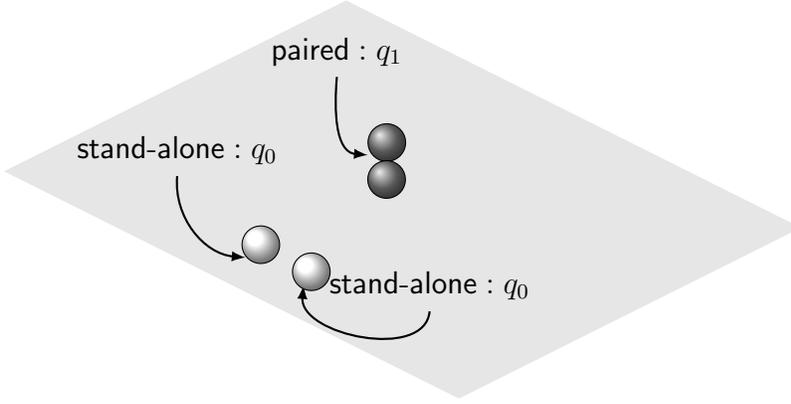

Let us for a moment consider on the $B$-Model, i.e. balls in stand-alone or paired state. First consider just two balls, as depicted in figure (\ref{fig:two.coins}), there are two configurations in the set $\mathcal{S}_2$. The normalisation condition in (\ref{eq:normalization_q_i}) reduces to
\begin{equation}
q_0 + q_1 = 1,
\end{equation}
and we have for the ratio
\begin{equation}
r = \frac{q_0}{q_1}, \qquad r \in [0, \infty),
\end{equation}
that
\begin{equation}
q_0 = \frac{r}{r + 1}, \qquad q_1 = \frac{1}{r + 1}.
\end{equation}
Since $r$ denote the ratio between the probabilities for not making a pair to making one,  $r < 1$ favours pairs to stand-alone states, $r = 1$ corresponds to equal probability and $r > 1$ favours stand-alone states to pairs.

For $n=3$, the number of possible pairs is the same as for two coins, in other words, $n_p \in \{0, 1\}$. Indeed, the difference between $n=3$ and $n=2$ is in the degenerate configurations for $n_p=1$: there are three distinguishable configurations, each with equal probability.

A similar relationship holds for all consecutive even and odd numbers, since the range of pair numbers $n_p \in \{0, 1, \dots, n\}$ is the same for $2n$ and $(2n+1)$. In fact $n = \floor{\frac{2n}{2}} = \floor{\frac{2n+1}{2}}$.
To stress the dependence of $q_i$ on the number of balls, namely $n$, we replace $q_i$ by $q_i^{(n)}$
\begin{equation}
P_{n}(n_p=i) = {n \choose 2i} (2i-1)!! \; q_{i}^{(n)}.
\end{equation}
Since $q_i^{(n)}$ is the probability for forming $i$ specific pairs amongst the $n$ balls, we have
\begin{equation}
q_i^{(n)} \propto r^{\floor{n/2}-i}.
\end{equation}
We normalise over the total number $W(n)$ of configurations by introducing a normalisation constant defined as
\begin{equation}
\label{eq:c_n_r_def}
c_n(r)  = \sum_{i=0}^{W(n)}  r ^{\floor{n/2}-i} = \sum_{k=0}^{\lfloor \frac{n}{2} \rfloor}  {n \choose 2k} (2k-1)!! \; r^{\lfloor \frac{n}{2} \rfloor - k},
\end{equation}
which allows us to write the probability for a specific state with $i$ pairs amongst $n$ balls as 
\begin{equation}
\label{eq:q_i_by_r}
  q_i^{(n)} = \frac{r^{\floor{n/2}-i}}{c_n(r)}.
\end{equation}

For the $B$-Model we arrive at the following expression for the probability distribution
\begin{equation}
\label{eq:p_n_n_p}
P_n(n_p=i) = \frac{1}{c_{n}(r)} {n \choose 2i} (2i-1)!! \; r^{\lfloor \frac{n}{2} \rfloor - i},
\end{equation}
whereas, for the $C$-Model we obtain a similar result that includes the head and tail states
\begin{equation}
\label{eq:p_n_n_p_n_h}
\fl
P_n(n_p=i, n_h = j)
 =\frac{1}{c_{n}(r)} {n \choose 2i} (2i-1)!! \; r^{\lfloor \frac{n}{2} \rfloor - i} \; {n-2i \choose j} \rho^j (1-\rho)^{n-2i-j} .
\end{equation}

\subsection{Finding the normalisation constant, $c_{n,r}$}

We now derive some properties of the normalisation constant, $c_{n}(r)$, that was introduced in (\ref{eq:c_n_r_def}).
\noindent In \ref{appendix:normalisation.constant.GF}, we obtain a recursive relation for even and odds normalization constants
\begin{equation}
\fl
\label{eq:both_c_2n_and_1}
c_{2n}(r)   = r c_{2n-1}(r) + (2n-1) c_{2n-2}(r), \qquad
c_{2n+1}(r) = c_{2n}(r)   + 2n c_{2n-1}(r).
\end{equation}
The last result for even numbers resembles the recursive rule of the state space volume in (\ref{eq:recursive}) despite a factor $r$ in place of $2$.

See table (\ref{table:c.n.r}) for the first eight iterations of $c_{n}(r)$ These are very similar to Hermite polynomials as function of $r$. Though all their coefficients are positive. \ref{appendix:hermit} represents the relation between Hermite polynomials and the normalization constant.

\begin{table}[H]
	\caption{\label{table:c.n.r}Normalization constant for first eight $n$s.}
	\begin{indented}
		\item[]\begin{tabular}{@{}ll}
			\br
			n & $c_{n}(r)$ \\
			\mr
			2 & $r + 1$ \\
			\hline
			3 & $r + 3$ \\
			\hline
			4 & $r^2 + 6r + 3$\\
			\hline
			5 & $r^2 + 10r + 15$\\
			\hline
			6 & $r^3 + 15r^2 + 45r + 15$ \\
			\hline
			7 & $r^3 + 21r^2 + 105r+ 105$ \\ 
			\hline
			8 & $r^4 + 28r^3 + 210r^2 + 420r + 105$ \\
			\hline
			9 & $r^4 + 36r^3 + 378r^2 + 1260r + 945$ \\
			\br
		\end{tabular}
	\end{indented}
\end{table}

We must stress that the new recursive relation is satisfied by the normalisation constant for an arbitrary, non-negative $r$. This result is more general than the state space volume recursive relation. It seems that the state space structure is not encoded only in a single recursive rule but also is valid for the normalisation constant of the probability distribution, parametrised by $r$ and $n$. 

In a statistical mechanics context, the normalisation constant $c_{2n}(r)$, is known as the partition function. Hence  Eq. (\ref{eq:both_c_2n_and_1}) describes the composition law of the partition function in the state space \cite{jensen2018, jeldtoft2018, Tempesta2020b}.

Also from the definition of $c_{2n}(r)$ in Eq. (\ref{eq:c_n_r_def}), the asymptotic leading terms for constant $r$ and $r \ll n$ is
\begin{equation}
\label{eq:c.2n.asymptotic}
c_{n}(r) \sim \frac{r^{\frac{1}{4}}}{2\rme} \left(\frac{n}{\rme}\right)^{\frac{n}{2}} \rme^{\sqrt{rn}} \left(1 + \Or(\frac{1}{\sqrt{n}})\right).
\end{equation}
The above asymptotic identity is derived directly from the result reported in \cite{jensen2018}.

Recall that $r \in [0, \infty)$, thus when $r$ is in the same order of $n$ or greater we obtain different asymptotic leading term for $c_{n}(r)$ like
\begin{equation}
\label{eq:c.2n.asymptotic.2}
c_{n}(r) \sim
\left\lbrace
\begin{array}{ll}
\frac{g(\frac{r}{n})^{-\frac{n}{2}g(\frac{r}{n})} \rme^{\sqrt{\frac{r n}{\rme}} }  (\frac{n}{\rme})^{n/2}}{ \sqrt{ \sqrt{\rme \epsilon}  f(\epsilon) g(\epsilon)}} & r \le \rme n\\
&\\
\frac{g(\frac{r}{n})^{-\frac{n}{2} g(\frac{r}{n})} r^{n/2}}{\sqrt{ \pi r f(\frac{r}{n}) g(\frac{r}{n})/2}} & r > \rme n
\end{array}
\right.,
\end{equation}
where
\begin{equation}
\label{eq:f_x_g_x}
f(x) = \sqrt{1 + \frac{4}{x}} - 1, \qquad g(x) = (1- \frac{x f(x)}{2}).
\end{equation}
Check \ref{appendix:asymptotic.leading.term.espsilon} for the details.

\section{Large deviation property}
\label{Large-devi}
The mathematical definition of the large deviation property (LDP) is represented in \ref{appendix:ldp}. It should suffice here to recall that a sequence of  probability measures $\{P_{n}; n = 1, 2, \dots\}$ is said to have a large deviation property if there exist a sequence of positive numbers $a_n$ (speed) for $ n \in \{1,2, \dots\}$ that tends to $\infty$, and a function $I(x)$ (rate function) such that the following limit exist
\begin{equation}
\lim_{n \rightarrow \infty} \frac{1}{a_n} \ln P_n(A) = -\inf_{x \in A} I(x).
\end{equation}

Denote the state of a ball in the $B$-model at index $l$, or a coin in the $C$-model, by a random variable $X_l$. Then for a particular configuration, random variables $n_p$ (the number of pairs) and $n_h$ (the number of heads) calculates as
\begin{equation}
n_p = \frac{1}{2} \sum_{l=0}^{n} \delta_{X_l, p}, \qquad n_h =  \sum_{l=0}^{n} \delta_{X_l, 1},
\end{equation}
where $\delta_{X_i, p}$ is the Kronecker delta and equal to one if $X_i$ is in a pair state, whereas $\delta_{X_i, 1}$ is equal to one for head states. In addition, the large deviation random variables of interest are defined as
\begin{equation}
\label{eq:m_{2n}_def}
m_{n} \equiv \frac{2n_p}{n} = \frac{1}{n} \sum_{i=0}^{n} \delta_{X_i, p}, \qquad 0 \le m_{n} \le 1,
\end{equation}
and 
\begin{equation}
\label{eq:s_{2n}_def}
s_{n} \equiv \frac{n_h}{n-2n_p} = \frac{1}{n-2n_p}  \sum_{i=0}^{n} \delta_{X_i, 1}, \qquad 0 \le s_{n} \le 1.
\end{equation}
In the continuum limit, $n \rightarrow \infty$, both variables are in $\mathbb{R}$ \cite{touchette2009}.

\ref{appendix:ldp.prob} finds the logarithm of $P_{n}(m_{n})$ in (\ref{eq:log_p_2n_complete}) as
\begin{equation}
\label{eq:log_p_2n_complete_text}
\fl
\ln P_n(m_{n}) = -\frac{1-m_{n}}{2}  (n \ln n)	
 - \frac{n}{2} \left[ m_{n}  \ln m_{n}  + (1 - m_{n}) \ln \frac{(1- m_{n})^2}{\rme r} \right] + \Or(\sqrt{n}).	
\end{equation}
It is interesting to see the resemblance of the terms inside the bracket and the Shannon entropy \cite{CoverThomasM2006}. For $0 \le p \le 1$, the Shannon entropy for a binary variable with probability $p$ is
\begin{equation}
H(p) = -p \ln p - (1-p) \ln (1-p),
\end{equation}
and along the same line, we define
\begin{equation}
\tilde{H_r}(p) \equiv - p \ln p  - (1 - p) \ln \frac{(1- p)^2}{\rme r},
\end{equation}
to rewrite the last result as
\begin{equation}
\label{eq:log_p_2n}
\ln P_n(m_{n}) =  -\frac{1-m_{n}}{2}  (n \ln n) + \frac{n}{2} \tilde{H_r}(m_{n}) + \Or(\sqrt{n}).
\end{equation}
Consequently, the large deviation limit is
\begin{equation}
\lim_{n \rightarrow \infty}  -\frac{1}{(n \ln n)} \ln P_n(m_{n}) = \frac{1-m_{n}}{2}.
\end{equation}
Notice that the speed is $a_n = n \ln n$. Hence, the large deviation probability of the $B$-model obtains as
\begin{equation}
    P_n(m_{n}) \asymp e^{-(n \ln n) I(m_{n})},
\end{equation}
for the rate function 
\begin{equation}
\label{eq:rate.f.1}
I(x) = \frac{1-x}{2}.
\end{equation}
When $n \ln n$ is not appreciably larger than $n$, we need to include the correction due to terms in $\Or(n)$ order as
\begin{equation}
I(x) = \frac{1-x}{2} - \frac{\tilde{H_r}(x) }{2 \ln n}, 
\end{equation}
or 
\begin{equation}
P(m_{n}) \asymp \exp\left[-(n \ln n)\frac{1-m_{n}}{2} + n\frac{\tilde{H_r}(m_{n})}{2}  \right].
\end{equation}

Although the large deviation rate function is $(1-m_{n})/2$, for practical purposes the $\Or(\frac{1}{\log n})$ correction must be included. For instance, let say the number of elements is in Avogadro's number order, or $n = 10^{23}$. Then $n \ln n = 53 \times 10^{23}$ is $53$ times larger than $n$, and $\tilde{H_r}(m_{n})/2$ can have comparable number of significant figures in comparison to $(1-m_{n})/2$.

For the $B$-model, \ref{appendix:ldp.prob} calculates (\ref{eq:log_p_2n_complete_2}) as
\begin{equation}
\label{eq:log_p_2n_complete_2_text}
\fl
\log P(m_{n}, s_{n}) =  -\frac{1-m_{n}}{2}  (n \ln n) - \frac{n}{2} \left[ 2(1-m_{n}) I_{\rho}(s_{n}) -\tilde{H_r}(m_{n}) \right] + \Or(\sqrt{n}),
\end{equation}
where $\rho$ is the probability of observing head state, and $I_{\rho}(s_{n})$ is the Bernoulli rate function \cite{touchette2009, Touchette2011}
\begin{equation}
I_{\rho}(x_2) = x_2 \ln (\frac{x_2}{\rho}) 
+  (1-x_2) \ln (\frac{1-x_2}{1-\rho}).
\end{equation}
Furthermore, the large deviation limit is
\begin{equation}
\lim_{n \rightarrow \infty}  -\frac{1}{(n \ln n)} \ln P(m_{n}, s_{n}) = \frac{1-m_{n}}{2},
\end{equation}
and 
\begin{equation}
   P_n(m_{n}, s_{n}) \asymp \exp\left[ -(n \ln n) I(m_{n}, s_{n}) \right]
\end{equation}
for the rate function
\begin{equation}
I(x_1, x_2) = 1-x_1 + \frac{1}{2 \ln n} \left[ 2(1-x_1) I_{\rho}(x_2) -\tilde{H_r}(x_1)  \right].
\end{equation}

\section{The probability distributions for the limiting case $\lim_{n \rightarrow \infty} \frac{r}{n} = \epsilon$}
\label{Limi-case}
For both $P_n(m_{n})$ and $P_n(m_{n}, s_{n})$ the minimum of the rate function is at $m_{n}=1$, independent of $r$ and $\rho$. In other words, in the thermodynamic limit ($n \rightarrow \infty$), all the balls or coins are in a paired state. Recall that the parameter $r$ is the relative abundance of non-pairs to pairs in both the $B$-Model and the $C$-Model. And seemingly, it does not affect the final outcome of distribution of states in the thermodynamic limit.

To explain this result, we refer to the degeneracy of the number of pairs in Sec. \ref{Bino-like}. In the thermodynamic limit, the terms ${n \choose 2n_p}(2n_p-1)!!$ grows fastest when $n_p = n/2$.  
So, using the definition of partitions of $\mathcal{S}$ or $\mathcal{S}'$ in Sec. \ref{Bino-like}, for any finite value of $r$, the cardinality of the subset $S_{i=n/2}$ is much larger than any other subsets, and their volumes are negligible in comparison to this single subset. Consequently in the limit, the probability of observing configurations in $S_{i=n/2}$ approaches to one.

As we shall see, in one special setting the $n \ln n$ speed is replaced by the usual linear $n$ speed.
Let us consider $r$ to be proportional to $n$. Recall $r \in [0, \infty)$, such that $r > 1$ assigns higher probability to stand-alone states. We will examine the large deviation probability of the same random variable when
\begin{equation}
0 <  \lim_{n \rightarrow \infty} \frac{r}{n} = \epsilon < \infty.
\end{equation}
This idea is similar to retrieving the Poisson distribution from the Binomial distribution, when $\rho n = \lambda$ is kept constant.

First we need to find the asymptotic leading term for $c_{n}(\epsilon)$, or the normalization constant in (\ref{eq:c.2n.asymptotic.2}). We do that by replacing $n \epsilon$ in place of $r$ (Note that (\ref{eq:c.2n.asymptotic}) was written for constant $r$ and is not applicable for this case). 
The asymptotic leading term of $\ln c_{n}(\epsilon)$ in (\ref{eq:c.2n.asymptotic.2}) finds as
\begin{equation}
\label{eq:log.c.2n.epsilon}
\ln c_{n}(\epsilon)
 =  \left\lbrace  \begin{array}{ll}
- \frac{n}{2} g(\epsilon) \ln g(\epsilon) + n\sqrt{\frac{\epsilon}{\rme}} +  \frac{n}{2}\ln \frac{n}{\rme}  + \Or(1)&, \epsilon \le \rme\\
- \frac{n}{2} g(\epsilon) \ln g(\epsilon) +  \frac{n}{2} \ln( \epsilon n) + \Or(\ln n)&, \epsilon > \rme
\end{array}\right.,
\end{equation}
where $g(\epsilon)$ is defined in (\ref{eq:f_x_g_x}). See \ref{appendix:asymptotic.leading.term.espsilon} for details.

Next, using $\ln c_{n}(\epsilon)$ and replacing $r$ by $\epsilon n$ for $\ln P_n(m_{n})$ in (\ref{eq:log_p_2n_complete_text}), the terms that are in order $\Or(n\ln n)$ cancels each others, and for the $B$-model it obtains
\begin{equation}
P_{\epsilon}(m_{n}) \asymp \rme^{ -n I_1(m_{n}; \epsilon)},
\end{equation}
where
\begin{equation}
I_1(x_1; \epsilon) = \frac{1}{2} \times \left\lbrace
\begin{array}{ll}
2\sqrt{\frac{\epsilon}{\rme}} - g(\epsilon) \ln g(\epsilon) -  \tilde{H_{\epsilon}}(x_1)&, 0 < \epsilon \le \rme \\
 \log(\epsilon \rme) -  g(\epsilon) \ln g(\epsilon) -   \tilde{H_{\epsilon}}(x_1)&, \epsilon > \rme
\end{array}
\right..
\end{equation}
Similarly for the $C$-model and $P_{\epsilon}(m_{n}, s_{n})$, (\ref{eq:log_p_2n_complete_2_text}) derives
\begin{equation}
P_{\epsilon}(m_{n}, s_{n}) \asymp \rme^{ -n I_2(m_{n}, s_{n}; \epsilon)},
\end{equation}
for
\begin{equation}
I_2(x_1, x_2; \epsilon) =  (1-x_1) I_{\rho}(x_2) +I_1(x_1; \epsilon).
\end{equation}
Notice that the logarithm of these two probability distributions are in order $O(n)$. 

\section{The closed form of $P_{n}(n_p)$ moments}
\label{Closed-form}
As we shall see, moments of $P_{n}(n_p)$, or the $B$-Model, can be expressed in closed form. Although the moments of the $C$-Model have the same feature, we do not report them here.

\subsection{First moment}
In \ref{appendix:moments}, in (\ref{eq:first.moment.even}) and (\ref{eq:first.moment.odd}), we find that the first moment of $P_n(n_p)$ and for even and odd numbers as
\begin{equation}
\langle n_p \rangle_{2n} =  n(2n-1)\frac{c_{2n-2}(r)}{c_{2n}(r)}, \qquad \langle n_p \rangle_{2n+1} = n(2n+1)\frac{c_{2n-1}(r)}{c_{2n+1}(r)}.
\end{equation}

Recall that $c_{n}(r)$ is the Hermite polynomial with positive coefficients, evaluated at $r$. And indeed, the first moment is proportional to the ratio of two Hermite polynomials.
For $m_{n} = 2n_p/n$, equation (\ref{eq:first.moment.m_n}) finds
\begin{equation}
\langle m_{n} \rangle_r = (n-1)\frac{c_{n-2}(r)}{c_{n}(r)}.
\end{equation}

For constant $r$ and $r \ll n$, the asymptotic expansion of $\langle m_{n} \rangle_r$ derives from the ratio of two asymptotic leading terms for $c_{n-2}(r)$ and $c_{n}(r)$. Using (\ref{eq:c.2n.asymptotic}), we get
\begin{eqnarray}
 \frac{c_{n-2}(r)}{c_{n}(r)} \sim \frac{\frac{r^{\frac{1}{4}}}{2\rme} \left(\frac{n}{\rme}\right)^{\frac{n}{2}-1} \left(1-\frac{2}{n}\right)^{\frac{n}{2}-1} \exp(\sqrt{rn-2r}) }{\frac{r^{\frac{1}{4}}}{2\rme} \left(\frac{n}{\rme}\right)^{\frac{n}{2}} \exp(\sqrt{rn}) } \sim \frac{\exp(-\sqrt{\frac{r}{n}})}{n},
\end{eqnarray}
and therefore
\begin{equation}
\label{eq:n_p_asymp}
\langle m_{n} \rangle_r \sim  \exp\left(-\sqrt{\frac{r}{n}}\right).
\end{equation}
Notice that $\langle m_{n} \rangle_r \rightarrow 1$ as $n \rightarrow \infty$. This is anticipated, as the minimum of the large deviation rate function in (\ref{eq:rate.f.1}) is $m_{n}=1$, independent of $r$. 

The second asymptotic leading term of $c_{n}(r)$ in (\ref{eq:log.c.2n.epsilon}), when $r \sim n$ or $r \gg n$, is given by a different asymptotic expression
\begin{equation}
\label{eq:first.momemt.both.regim}
\langle m_{n} \rangle \sim 
	\left\lbrace
	\begin{array}{ll}
	\exp\left(-\sqrt{\frac{r}{n}}\right)&, r \le \rme n  \\
	g(\frac{r}{n})^{g(\frac{r}{n})} \frac{n}{r}&, r > \rme n
	\end{array}.
\right.
\end{equation} 
The numerical study of the above result shows that its estimate is more accurate for values of $r$ that are not close to $\rme n$.

And finally, the large deviation expectation of $m_{n}$ with respect to $P_{\epsilon}(m_{n})$, or  for the regime that $r/2n = \epsilon$ is kept constant, is
\begin{equation}
\langle m_{n} \rangle_{\epsilon} \sim 
\left\lbrace
	\begin{array}{ll}
	\exp\left(-\sqrt{\epsilon}\right)&, \epsilon \le \rme  \\
	\frac{g(\epsilon)^{g(\epsilon)}}{\epsilon}&, \epsilon > \rme 
	\end{array}
\right..
\end{equation} 

\subsection{Other moments}
\ref{appendix:moments} finds the $k$-th moment in (\ref{eq:k.moments}) with respect to $P_{n}(n_p)$ as
\begin{equation}
\langle n_{p}^{k} \rangle_{n} = \sum_{i=1}^{k} a^{(k)}_{i} \frac{\floor{\frac{n}{2}}! (n-1)!!}{(\floor{\frac{n}{2}}-i)!(n-2i-1)!!} \frac{c_{n-2i}(r)}{c_{n}(r)},
\end{equation}
where the recursive relation for $a^{(k)}_{i}$ is
\begin{equation}
\label{eq:a.k.i.rec}
a^{(k)}_{i} = a^{(k-1)}_{i-1} + ia^{(k-1)}_{i}, \qquad a^{(k)}_{1} = a^{(k)}_{k} = 1.
\end{equation}
Check \ref{appendix:moments} for details.

\subsection{A relation among first moments with different sizes}
In this part, we show an identity that relates the $k$-th moment of a system with length $n$ to the first moment of systems with smaller sizes. In \ref{appendix:moments.sizes.relation}, it is shown in (\ref{eq:k.moment.to.smaller.size}) that
\begin{equation}
\langle n_{p}^{k} \rangle_{n} = \sum_{i=1}^{k} a^{(k)}_{i} \langle n_{p} \rangle_{n} \langle n_{p} \rangle_{(n-2)} \langle n_{p} \rangle_{(n-4)} \dots \langle n_{p} \rangle_{(n-2i+2)}, 
\end{equation}
where $a^{(k)}_{i}$ is defined in (\ref{eq:a.k.i.rec}). Note that all the summand terms are the first moment for smaller system sizes. \textit{e.g.} the average $\langle n_{p} \rangle_{(k)}$ is taken with respect to $P_{k}(n_p)$.
To find the variance of the random variable $n_p$ with respect to $P_{n}(n_p)$, the second moment writes as
\begin{equation}
\langle n_{p}^{2} \rangle_{n} = a^{(2)}_{1} \langle n_{p} \rangle_{n} + a^{(2)}_{2} \langle n_{p}\rangle_{n} \langle n_{p} \rangle_{(n-2)}
= \langle n_{p}\rangle_{n} \left[1 + \langle n_{p}\rangle_{(n-2)} \right].
\end{equation}
Hence,
\begin{equation}
\Var[n_p]_{n} = \langle n_p^2\rangle_{n} - \langle n_p \rangle^2_{n}
= \langle n_{p} \rangle_{n} \left[1 + \langle n_{p} \rangle_{(n-2)}  - \langle n_{p} \rangle_{n} \right].
\end{equation}

\subsection{Probability generating function}
\ref{appendix:generating.func} finds the probability generating function for random variables $n_p$ and $n_h$. Recall that the normalization constant $c_{n}(r)$ is a Hermite polynomial degree $n$, evaluates at $r$.
For $n_p$, the probability generating function is
\begin{equation}
G_{n}(s) = s^{\frac{n}{2}} \frac{c_{n}(\frac{r}{s})}{c_{n}(r)},
\end{equation}
where $c_{n}(\frac{r}{s})$ is the normalization constant, evaluates at $\frac{r}{s}$.

Similarly, for the probability generating function for random variables $n_p$ and $n_h$ is
\begin{equation}
G_{n}(s, u) = s^{\frac{n}{2}} \frac{c_{n}\left(\frac{r (\rho u + 1 - \rho )^{2}}{s}\right)}{c_{n}(r)}.
\end{equation}
The normalization constant in the numerator evaluates at $\left(\frac{r \left(\rho u + 1 - \rho \right)^{2}}{s}\right)$.

\section{Marginal distributions of $P_{n}(n_p)$ and $P_{n}(n_p, n_h)$}
\label{Marginal-dist}

The $B$-model marginal distribution is defined for a single random variable, say $X_l$, as a function that projects the state of a ball at $l$ to its domain
\begin{equation}
X_l: \mathcal{S}_{n} \rightarrow \{0, 1\},
\end{equation}
where $\mathcal{S}_{n}$ is the state space of the $B$-Model. $X_l=0$ corresponds to a paired state and $X_l=1$ to a stand-alone ball.

For the $C$-Model, in the case of single coin in head or tail state
\begin{equation}
X_l: \mathcal{S'}_{n} \rightarrow \{-1, 0, 1\},
\end{equation}
for state space $\mathcal{S'}_{n}$, $X_l=1$ for the head, $X_l=0$ for the paired state, and $X_l=-1$ for the tail state.

To find the marginal distribution, summing the probabilities over configurations that $X_l$ has the same value suffices
\begin{equation}
\label{eq:marginal_main_sum}
P_{n}(X_l = x) = \sum_{c \in S_{n}: X_l(c) = x} P_{n}(c).
\end{equation}
Therefore, partitioning the state space $\mathcal{S}_{n}$ or $\mathcal{S}'_{n}$ according to the state of an element at $l$ is the key to find the marginals. Intuitively, we understand that a marginal is invariant with respect to $l$, or say, identical for all $l$s. This means that $l$ is an arbitrary site.

\subsection{Marginal distribution of $P_{n}(n_p)$}
\begin{figure}[h]
	\centering
	\includegraphics[width=0.49\textwidth]{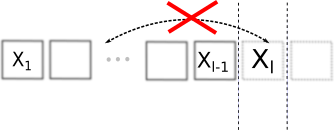}   
	\caption{An example of configurations in subset $S_{i}^{(1)}$.} 
	\label{fig:ts_division_1}	
\end{figure}

\begin{figure}[h]
	\centering
	\includegraphics[width=0.49\textwidth]{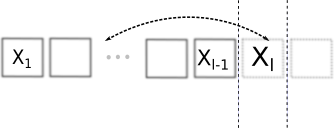}  
	\caption{An example of configurations in subset $S_{i}^{(2)}$.} 
	\label{fig:ts_division_2}	
\end{figure}

Let us start with the $B$-Model. In Sec. \ref{Bino-like}, $S_i$ introduced as the subsets of $\mathcal{S}_{n}$ that has $i$ pairs. Next, partition the elements of $S_i$ to two disjoint subsets as
\begin{equation}
S_i = S_{i}^{(1)} \bigcup S_{i}^{(2)}, \qquad S_{i}^{(1)} \bigcap S_{i}^{(2)} = \emptyset,
\end{equation}
such that $S_{i}^{(1)}$ contains only the configurations that do not have a pair link to site $l$ --- Figure (\ref{fig:ts_division_1})--- whereas $S_{i}^{(2)}$ contains those configurations that have one ---  Figure (\ref{fig:ts_division_2}).

In other words, for every configuration in $S_{i}^{(1)}$, the ball at $l$ is in a stand-alone state, whilst for configurations belong to $S_{i}^{(2)}$, the ball at $l$ are in a paired state. 

After that, we partition $S_{i}^{(2)}$ to disjoint subsets
\begin{equation}
S_{i}^{(2)} = \bigcup_{\begin{array}{l}
	k=1 \\ 
	k \ne l
	\end{array}}^{n} S_{i,k}^{(2)}, \qquad S_{i,k}^{(2)} \bigcap S_{i,h}^{(2)} = \emptyset \quad k \ne h,
\end{equation}
such that, $S_{i,k}^{(2)}$ contains the configurations that has a paired state between site $k$ and $l$. 
The partitioning of $\mathcal{S}_{n}$ allows us to rewrite the marginal sum in (\ref{eq:marginal_main_sum}) as
\begin{equation}
\fl
P_{n}\{ \mbox{the site } l \mbox{ in a stand-alone state} \} = P_{n}(X_l = 1)  = \sum_{i=0}^{n/2} P(S_{i}^{(1)}),
\end{equation}
and
\begin{equation}
\fl
P_{n}\{ \mbox{the site } l \mbox{ in a paired state} \}
 = P_{n}(X_l = 0)  = \sum_{i=0}^{n/2} \sum_{\begin{array}{l}
	k=1 \\ 
	k \ne l
	\end{array}}^{n} P(S_{i,k}^{(2)}).
\end{equation}
From (\ref{eq:q_i_equive_to_P_c_c_i}) and (\ref{eq:q_i_by_r}), the probability of a observing a single configurations $c_i \in S_{i}$ is equal to
\begin{equation}
P_{n}(c_i) = \frac{r^{\frac{n}{2}-i}}{c_{n}(r)},
\end{equation} 
so, we need to find the cardinality of $S_{i}^{(1)}$ and $S_{i,k}^{(2)}$ to find $P(S_{i}^{(1)})$ and $P(S_{i,k}^{(2)})$, respectively.

Form its definition, $S_{i}^{(1)}$ contains $i$ pairs while the site $l$ is in a stand-alone state. Indeed, we need to choose $2i$ paired balls among $n-1$ candidates, and there are $(2i-1)!!$ distinguishable permutations among $i$ pairs. It concludes that the cardinality of $S_{i}^{(1)}$ must be
\begin{equation}
\label{eq:s_i_cardinality}
|S_{i}^{(1)}| = {n-1 \choose 2i} (2i-1)!!.
\end{equation}
For $S_{i,k}^{(2)}$, one of the pairs has already been selected, so there are $2i-2$ choices from $n-2$ candidates, and $(2i-3)!!$ permutations among them. It results in
\begin{equation}
\label{eq:s_i_k_cardinality}
|S_{i,k}^{(2)}| = {n-2 \choose 2i-2} (2i-3)!!.
\end{equation}
\ref{appendix:marginal.sums} shows the details of the calculation, and finds
\begin{equation}
P_{n}(x) = 
\left\lbrace
\begin{array}{ll}
\frac{2\langle n_p \rangle}{n}&, x = 0  \\
1 - \frac{2\langle n_p \rangle}{n}&, x = 1 
\end{array}
\right.
= 
\left\lbrace
\begin{array}{ll}
(n-1) \frac{c_{n-2}(r)}{c_{n}(r)}&, x = 0  \\
1 - (n-1) \frac{c_{n-2}(r)}{c_{n}(r)}&, x = 1 
\end{array}
\right..
\end{equation}
To find the large deviation marginal, or for $n \gg 1$, the normalization constant in (\ref{eq:first.momemt.both.regim}) for $r < \rme n$ results in 
\begin{equation}
P_{n}(x) = 
\left\lbrace
\begin{array}{ll}
\exp\left(-\sqrt{\frac{r}{n}}\right)&, x = 0  \\
1 - \exp\left(-\sqrt{\frac{r}{n}}\right)&, x = 1 
\end{array}
\right.,
\end{equation}
and for $r \ge \rme n$
\begin{equation}
P_{n}(x) = 
\left\lbrace
\begin{array}{ll}
g(\frac{r}{n})^{g(\frac{r}{n})} \frac{n}{r}&, x = 0  \\
1 - g(\frac{r}{n})^{g(\frac{r}{n})} \frac{n}{r}&, x = 1 
\end{array}
\right..
\end{equation}
\subsection{Marginal distribution of $P_{n}(n_p, n_h)$}
When we apply the same argument for $P_{n}(n_p, n_h)$ similar to the previous part, we must sum over head and tail states too. Doing that we get
\begin{equation}
P_{n}(x) = 
\left\lbrace
\begin{array}{ll}
(1-\rho) \left(1 - (n-1) \frac{c_{n-2}(r)}{c_{n}(r)}\right)&, x = -1  \\
(n-1) \frac{c_{n-2}(r)}{c_{n}(r)}&, x = 0  \\
\rho \left(1 - (n-1) \frac{c_{n-2}(r)}{c_{n}(r)}\right)&, x = 1 
\end{array} 
\right..
\end{equation}
The large deviation marginal for $n \gg 1$ and $r < \rme n$ is
\begin{equation}
P_{n}(x) = 
\left\lbrace
\begin{array}{ll}
(1-\rho) \left(1 - \exp\left(-\sqrt{\frac{r}{n}}\right)\right) &, x = -1  \\
\exp\left(-\sqrt{\frac{r}{n}}\right)&, x = 0  \\
\rho \left(1 - \exp\left(-\sqrt{\frac{r}{n}}\right)\right)&, x = 1 
\end{array}
\right.,
\end{equation}
while for $r \ge \rme n$ is
\begin{equation}
P_{n}(x) = 
\left\lbrace
\begin{array}{ll}
(1-\rho) \left(1 - g(\frac{r}{n})^{g(\frac{r}{n})}\right) &, x = -1  \\
g(\frac{r}{n})^{g(\frac{r}{n})}&, x = 0  \\
\rho \left(1 - g(\frac{r}{n})^{g(\frac{r}{n})}\right)&, x = 1 
\end{array}
\right..
\end{equation}

\section{Summary}
We have analysed the statistics of two simple models by use of combinatorial arguments. One consisting of balls ($B$-Model) with no internal structure and another consisting of coins ($C$-model) with single particle states: head or tail. In both cases new emergent pair states can form .

The simplicity of the models permits a detailed analysis of the probability distribution $P_n(n_p)$ for observing $n_p$ pairs in the $B$-model and the probability distribution $P_n(n_p, n_h)$ for observing $n_p$ pairs together with $n_h$ heads in the $C$-model.

Both probability distribution satisfy large deviation principles with speed equal to $n \ln n$. We derived order $O(n)$  correction terms which will make important contributions to the leading $O(n \ln n)$ terms except in the ultimate asymptotic limit. 

When pairing is gradually frozen out by tuning the ratio $r = q_0/q_1$ between the non-pairing ($q_0$) and the pairing ($q_1$) probabilities such that $r$ decreases inversely proportional to the increasing number of balls, i.e.  $n$, the probability distribution for pairs satisfies a large deviation principle with the usual speed $n$.

Both the original coin model introduced in \cite{jensen2018} and the new ball model introduce in the present paper can be seen as minimalistic and paradigmatic models for statistical analysis of systems with emergent structures. For example, the $C$-Model can be viewed as an Ising model where pairs of Ising spins can combine to form new states.

For reference we  have included an appendix \ref{appendix:tables} with tables listing statistical properties, which may be useful as reference points for studies of more involved models of state spaces with new emergent many-component states.

\appendix

\section{Tables}
\label{appendix:tables}

\begin{table}[H]
	\caption{\label{table:p_2n_n_p}The probability distribution for pairs and non-paired states.}
	\begin{indented}
		\item[]\begin{tabular}{@{}ll}
			\br
			Name & Mathematical form \\
			\br
			Probability Distribution & $P_{n}(n_p = i) = \frac{{n \choose 2i} (2i-1)!! \; r^{\floor{\frac{n}{2}}-i}}{c_{n}(r)}$  \\			
			\mr
			Normalization constant& $c_{n}(r) = \sum_{k=0}^{\floor{n/2}}  {n \choose 2k} (2k-1)!! \; r^{\floor{\frac{n}{2}} - k}$\\
			& (Hermite Polynomial with positive coefficients)\\
			\mr
			First Moment& $\langle n_p \rangle_{2n} = n(2n-1)\frac{c_{2n-2}(r)}{c_{2n}(r)}$ \\
			& $\langle n_p \rangle_{2n+1} = n(2n+1)\frac{c_{2n-1}(r)}{c_{2n+1}(r)}$\\
			& $\langle m_n \rangle_{r} = (n-1)\frac{c_{n-2}(r)}{c_{n}(r)}$\\
			& $m_{n} = \frac{1}{n} \sum_{i=0}^{n} \delta_{X_i, p}$ \\
			\mr
			Other Moments (even $n$)& $\langle n_{p}^{k} \rangle_{n} = \sum_{i=1}^{k} a^{(k)}_{i} \frac{\floor{\frac{n}{2}}! (n-1)!!}{(\floor{\frac{n}{2}}-i)!(n-2i-1)!!} \frac{c_{n-2i}(r)}{c_{n}(r)}$ \\
			&$a^{(k)}_{i} = a^{(k-1)}_{i-1} + ia^{(k-1)}_{i}$\\
			&$a^{(k)}_{1} = a^{(k)}_{k} = 1$\\
			\mr 
			Marginal& $P_{n}(X_l) = 
			\left\lbrace
			\begin{array}{ll}
			(n-1) \frac{c_{n-2}(r)}{c_{n}(r)}&, X_l = 0  \\
			1 - (n-1) \frac{c_{n-2}(r)}{c_{n}(r)}&, X_l = 1 
			\end{array}
			\right.$\\
			\mr 
			Large deviation form & $P_n(m_{n}) \asymp \exp\left[-(n \ln n) (\frac{1-m_{n}}{2})\right]$\\
			& $m_{n} = \frac{1}{n} \sum_{i=0}^{n} \delta_{X_i, p}$\\
			\mr 
			Large deviation & $P_n(m_{n}) \asymp \exp\left[-(n \ln n) (\frac{1-m_{n}}{2}) + \frac{n}{2} \tilde{H_r}(m_{n})\right]$\\
			(second order) & $\tilde{H_r}(x):= - x \ln x  - (1 - x) \ln \frac{(1- x)^2}{\rme r}$ \\
			\mr 
			Large deviation first moment & $\langle m_{n} \rangle_{r} = \exp\left(-\sqrt{\frac{r}{n}}\right)$ \\
			\mr 
			Marginal ($n \gg 1$) & $P_{n}(x) = 
			\left\lbrace
			\begin{array}{ll}
			\exp\left(-\sqrt{\frac{r}{n}}\right)&, x = 0  \\
			1 - \exp\left(-\sqrt{\frac{r}{n}}\right)&, x = 1 
			\end{array}
			\right.$ \\
			\br
		\end{tabular}
	\end{indented}
\end{table}

\begin{table}[H]
	\caption{\label{table:p_2n_n_p_n_h}The probability distribution including head and tail states.}
	\begin{indented}
		\item[]
		\begin{tabular}{@{}ll}
			\br
			Name & Mathematical form \\
			\br
			Probability Distribution & $P_{n}(n_p = i, n_h = j) = \frac{{n \choose 2i} (2i-1)!! \; r^{\floor{\frac{n}{2}}-i} }{c_{n}(r)} \; {n-2i \choose j} \rho^j (1-\rho)^{n-2i-j}$  \\			
			\mr
			Normalization constant& $c_{n}(r) = \sum_{k=0}^{\floor{n/2}}  {n \choose 2k} (2k-1)!! \; r^{\floor{\frac{n}{2}} - k}$\\
			& (Hermite Polynomial with positive coefficients)\\
			\mr 
			Marginal& $P_{n}(X_l) =
			\left\lbrace
			\begin{array}{ll}
			(1-\rho) \left(1 - (n-1) \frac{c_{n-2}(r)}{c_{n}(r)}\right)&, X_l = -1 \\
			(n-1) \frac{c_{n-2}(r)}{c_{n}(r)}&, X_l = 0 \\
			\rho \left(1 - (n-1) \frac{c_{n-2}(r)}{c_{n}(r)}\right)&, X_l = 1
			\end{array}
			\right.$\\
			\mr 
			Large deviation form & $P_n(m_{n}, s_{n}) \asymp \exp\left[-(n \ln n) (\frac{1-m_{n}}{2})\right]$\\
			& $m_{n} = \frac{1}{n} \sum_{i=0}^{n} \delta_{X_i, p}$\\
			& $s_{n} = \frac{1}{n-2n_p}  \sum_{i=0}^{n} \delta_{X_i, 1}$ \\
			\mr 
			Large deviation & $P_n(m_{n}, s_{n}) \asymp \exp\left[-(n \ln n) (\frac{1-m_{n}}{2}) - n (1-m_{n}) I_{\rho}(s_{n}) + \frac{n}{2} \tilde{H_r}(m_{n})\right]$\\
			(second order) & $I_{\rho}(x):= x \ln (\frac{x}{\rho}) 
			+  (1-x) \ln (\frac{1-x}{1-\rho})$ \\
			& $\tilde{H_r}(x):= - x \ln x  - (1 - x) \ln \frac{(1- x)^2}{\rme r}$ \\
			\mr 
			Marginal ($n \gg 1$) & $P_{n}(x) = 
			\left\lbrace
			\begin{array}{ll}
			(1-\rho) \left(1 - \exp\left(-\sqrt{\frac{r}{n}}\right)\right)&, x = -1 \\
			\exp\left(-\sqrt{\frac{r}{n}}\right)&, x = 0 \\
			\rho \left(1 - \exp\left(-\sqrt{\frac{r}{n}}\right)\right)&, x = 1
			\end{array}
			\right.$ \\
			\br
		\end{tabular}
	\end{indented}
\end{table}

\begin{table}[H]
	\caption{\label{table:p_2n_epsilon}The probability distribution $P_{\epsilon}(m_{n})$.}
	\begin{indented}
		\item[]\begin{tabular}{@{}ll}
			\br
			Name & Mathematical form \\
			\br
			Large deviation form & $P_{\epsilon}(m_{n}) \asymp \exp\left[-n I_1(m_{n}; \epsilon)\right]$\\ 
			& $m_{n} = \frac{1}{n} \sum_{i=0}^{n} \delta_{X_i, p}$\\ 
			& $I_1(x_1; \epsilon) = \frac{1}{2} \times \left\lbrace
			\begin{array}{ll}
			2\sqrt{\frac{\epsilon}{\rme}} - g(\epsilon) \ln g(\epsilon) -  \tilde{H_{\epsilon}}(x_1)&, 0 < \epsilon \le \rme \\
			\log(\epsilon \rme) -  g(\epsilon) \ln g(\epsilon) -   \tilde{H_{\epsilon}}(x_1)&, \epsilon > \rme
			\end{array}
			\right.$\\
			& $g(\epsilon) = 1- \frac{\epsilon }{2}\left(\sqrt{1 + \frac{4}{\epsilon}} - 1 \right)$\\		
			\mr 
			First moment & $\langle m_{n} \rangle_{\epsilon} \sim
			\left\lbrace
			\begin{array}{ll}
			\exp\left(-\sqrt{\epsilon}\right)&, \epsilon \le \rme  \\    
			\frac{g(\epsilon)^{g(\epsilon)}}{\epsilon}&, \epsilon > \rme  
			\end{array}
			\right.
			$ \\
			\mr 
			Marginal & $P_{\epsilon}(x) = 
			\left\lbrace
			\begin{array}{ll}
			\exp\left(-\sqrt{\epsilon}\right)&,  x = 0  \\
			1 - \exp\left(-\sqrt{\epsilon}\right)&, x = 1 
			\end{array}
			\right.$ \\
			\br
		\end{tabular}
	\end{indented}
\end{table}

\begin{table}[H]
	\caption{\label{table:p_2n_epsilon_2}The probability distribution $P_{\epsilon}(m_{n}, s_{n})$.}
	\begin{indented}
		\item[]\begin{tabular}{@{}ll}
			\br
			Name & Mathematical form \\
			\br
			Large deviation form & $P_{\epsilon}(m_{n}, s_{n}) \asymp \exp\left[ -n I_2(m_{n}, s_{n}; \epsilon)\right]$\\  
			& $m_{n} = \frac{1}{n} \sum_{i=0}^{n} \delta_{X_i, p}$\\
			& $s_{n} = \frac{1}{n-2n_p}  \sum_{i=0}^{n} \delta_{X_i, 1}$ \\	
			& $I_2(x_1, x_2; \epsilon) =  (1-x_1) I_{\rho}(x_2) +I_1(x_1; \epsilon)$\\			
			&$I_{\rho}(x):= x \ln (\frac{x}{\rho}) 
			+  (1-x) \ln (\frac{1-x}{1-\rho})$\\							
			\mr 
			Marginal & $P_{\epsilon}(x) = 
			\left\lbrace
			\begin{array}{ll}
			(1-\rho) \left[1 - \exp\left(-\sqrt{\epsilon}\right)\right] &, x = -1  \\
			\exp\left(-\sqrt{\epsilon}\right)&, x = 0  \\
			\rho \left[1 - \exp\left(-\sqrt{\epsilon}\right)\right]&, x = 1 
			\end{array}
			\right.$ \\
			\br
		\end{tabular}
	\end{indented}
\end{table}

\section{The Hermit polynomials and normalization constant}
\label{appendix:hermit}
Recall that the Hermit polynomials define as
\begin{equation}
  H_{n}(x) = (-1)^{n} \rme^{\frac{x^2}{2}} \frac{d^n}{dx^n} \rme^{-\frac{x^2}{2}}.
\end{equation}
We define Hermit polynomials with positive coefficients, say $H^{(+)}_{n}(x)$ as 
\begin{equation}
H^{(+)}_{n}(x) = \rme^{-\frac{x^2}{2}} \frac{d^n}{dx^n} \rme^{\frac{x^2}{2}}.
\end{equation}
Then, for even sizes, namely $2n$, the normalization constant defines 
\begin{equation}
   c_{2n}(r) = H^{(+)}_{2n}(\sqrt{r}),
\end{equation} 
and for odd sizes one has
\begin{equation}
c_{2n+1}(r) = \frac{H^{(+)}_{2n+1}(\sqrt{r})}{\sqrt{r}}.
\end{equation}

\section{The large deviation property}
\label{appendix:ldp}
In \cite{Ellis85}, \textbf{Definition II.3.1}, the large deviation property (LDP) defines as follows:\\

\noindent Let $\Omega$ be a complete separable metric space, and $\mathcal{B}(\Omega)$ the Borel $\sigma$-field of $\Omega$. 
A sequence of probability measures $\{P_{n}; n = 1, 2, \dots\}$ is said to have a large deviation property if there exist a sequence of positive numbers $\{a_n; n= 1,2, \dots\}$ (speed) that tend to $\infty$, and a function $I(x)$ (rate function) that maps $\Omega$ into $[0, \infty]$ such that the followings hold:
\begin{enumerate}
	\item $I(x)$ is lower semicontinuous on $\Omega$ and has compact level sets.
	\item $\lim \sup_{n \rightarrow \infty} \ln P_n(K)/a_n \le -\inf_{x \in K}I(x)$ for each closed set $K$ in $\Omega$.
	\item $\lim \inf_{n \rightarrow \infty} \ln P_n(G)/a_n \ge -\inf_{x \in G}I(x)$ for each open set $G$ in $\Omega$.
\end{enumerate}

Then any Borel subset of $\Omega$, say, $A$, that is an \textit{I-continuity set} has the limit \cite{Ellis85} 
\begin{equation}
\lim_{n \rightarrow \infty} \frac{1}{a_n} \ln P_n(A) = -\inf_{x \in A} I(x).
\end{equation}
The set $A$ is an I-continuity if 
\begin{equation}
\inf_{x \in cl(A)} I(x) = \inf_{x \in in(A)} I(x),
\end{equation}
where $cl(A)$ and $in(A)$ denote closure and interior of $A$ respectively.

\section{Finding the normalization constant's recursive relation}
\label{appendix:normalisation.constant.GF}
To find the normalization constant $c_{n}(r)$, defined in (\ref{eq:c_n_r_def}), we introduce a generating function as
\begin{equation}
\label{eq:f.2n.x}
f_{2n}(x) = \sum_{i=0}^{n}  {2n \choose 2i} (2i-1)!! \; x^i.
\end{equation}
Then, for both even and odd numbers it rewrites the normalization constants as
\begin{equation}
\label{eq:c_2n_both_cases}
c_{2n}(r) = r^n f_{2n}(\frac{1}{r}), \qquad c_{2n+1}(r) = r^n f_{2n+1}(\frac{1}{r}).
\end{equation}
Afterwards, summing both sides of (\ref{eq:f.2n.x}) constructs $F_1(Y; x)$ as a generating function that its $Y^{2n}$ coefficient is equal to $f_{2n}(x)$
\[
F_1(Y; x) = \sum_{2n \ge 0} f_{2n}(x) \frac{Y^{2n}}{2n!}  = \sum_{2n \ge 0} \sum_{i=0}^{n} \frac{(\frac{x}{2})^i}{i!} \frac{Y^{2n}}{(2n-2i)!}
\]
\begin{equation}
=  \sum_{i \ge 0} \frac{(\frac{x}{2}Y^2)^i}{i!} \sum_{2n \ge 0} \frac{Y^{2n}}{2n!}
=  \exp(\frac{x}{2}Y^2) \cosh Y.
\end{equation}
Similarly, for odd numbers
\begin{equation}
F_2(Y; x) = \sum_{2n+1 \ge 1} f_{2n+1}(x) \frac{Y^{2n+1}}{(2n+1)!} = \exp(\frac{x}{2}Y^2) \sinh Y.
\end{equation}
Moreover, adding $F_1(Y; x)$ and $F_2(Y; x)$ together results in a generating function for even and odd powers  
\[
F(Y, x) \equiv  \sum_{2n \ge 0} f_{2n}(x) \frac{Y^{2n}}{2n!} + \sum_{2n+1 \ge 1} f_{2n+1}(x) \frac{Y^{2n+1}}{(2n+1)!}
\]
\begin{equation}
\label{eq:gen.function.F.Y.x}
= \sum_{n \ge 0} f_{n}(x) \frac{Y^{n}}{n!} = F_1(Y; x) + F_2(Y; x) = \exp(\frac{x}{2}Y^2 + Y).
\end{equation}

On summing $F_1(Y; x)$ and $F_2(Y; x)$, we wrote $F(Y, x)$ as a power series with odd and even powers of $Y$. $F(Y, x)$ is an analytic function and infinitely many differentiable. Hence, the coefficients of the Taylor expansion of $F(Y, x)$ at $Y=0$ obtains $f_{n}(x)$
\begin{equation}
\label{eq:fn_x.by.derivative}
f_{n}(x) = \left . \frac{\rmd^n F(Y, x)}{\rmd Y^n} \right|_{Y=0} \equiv F^{(n)}(0, x).
\end{equation}
Since the first derivative of $F(Y,x) = \exp\left[(xY^2)/2 + Y\right]$ is recursively relates to itself
\begin{equation}
F'(Y,x) = (xY+1)F(Y,x) \Rightarrow F'(0,x)  = 1,
\end{equation}
repetitively taking the derivatives finds the recursive equation for $F^{(n+1)}(Y, x)$ as
\begin{equation}
F^{(n+1)}(Y,x) = (xY+1)F^{(n)}(Y,x) + nxF^{(n-1)}(Y,x).
\end{equation}
And for $Y=0$
\begin{equation}
F^{(n+1)}(0,x) = F^{(n)}(0,x) + nxF^{(n-1)}(0,x).
\end{equation}
Plugging in (\ref{eq:fn_x.by.derivative}) into the above equation finds
\begin{equation}
\label{eq:f_n_1__f_n_n_x_f_n_1}
f_{n+1}(x) = f_{n}(x) + nxf_{n-1}(x).
\end{equation}
It is necessary to rewrite $f_{n+1}(x)$ for odds and even numbers separately, as it showed in (\ref{eq:c_2n_both_cases}). It means that the governing recursive equations are
\begin{equation}
\fl
c_{2n}(r)   = r c_{2n-1}(r) + (2n-1) c_{2n-2}(r), \qquad
c_{2n+1}(r) = c_{2n}(r)   + 2n c_{2n-1}(r).
\end{equation}

\section{Large deviation probabilities}
\label{appendix:ldp.prob}
Let us start with $P_{n}(n_p)$ in (\ref{eq:p_n_n_p}). Using Sterling's approximation for $\ln (n!)$, we get
\[
\fl
\ln P_{n}(n_p)  = - \ln c_{n}(r) + n_p \ln n -\frac{n}{2} \left[ \frac{2n_p}{n} \ln \frac{2n_p}{n} + 2(1 - \frac{2n_p}{n}) \ln (1- \frac{2n_p}{n}) \right.
\]
\begin{equation}
\label{eq:log_p_2n_n_p}
\left.   + \frac{2n_p}{n}  - (1 - \frac{n_p}{n}) \ln r \right].  	
\end{equation}
Equation (\ref{eq:m_{2n}_def}) defines the random variable $m_{n} = 2n_p/n$. Therefore, the last equation can be written in terms of $m_{n}$ as
\[
\fl
\ln P_{n}(m_{n})  = - \ln c_{n}(r) + \frac{m_{n}}{2} (n \ln n) -\frac{n}{2} \left[ m_{n} \ln m_{n} + 2(1 - m_{n}) \ln (1- m_{n}) \right.
\]
\begin{equation}
\left.   + m_{n}  - (1 - m_{n}) \ln r \right].  	
\end{equation}
After using the asymptotic leading term of $c_{n}(r)$ in (\ref{eq:c.2n.asymptotic}), it finds
\begin{equation}
\label{eq:log_p_2n_complete}
\fl
\ln P_n(m_{n}) = -\frac{1-m_{n}}{2}  (n \ln n)	
- \frac{n}{2} \left[ m_{n}  \ln m_{n}  + (1 - m_{n}) \ln \frac{(1- m_{n})^2}{\rme r} \right] + \Or(\sqrt{n}).	
\end{equation}
When one divides $\ln P_n(m_{n})$ by $-(n \ln n)$, as $n \rightarrow \infty$ the LDP limit exists
\begin{equation}
\label{eq:ldp.lim.p_m_n}
\lim_{n \rightarrow \infty}  -\frac{1}{(n \ln n)} \ln P_n(m_{n}) = \frac{1-m_{n}}{2}.
\end{equation}
\noindent\\ Repeating the same for $P_n(m_{n}, s_{n})$ in (\ref{eq:p_n_n_p_n_h}), one finds
\begin{equation}
\label{eq:log_p_2n_complete_2}
\fl
\log P_n(m_{n}, s_{n}) =  -\frac{1-m_{n}}{2}  (n \ln n) - \frac{n}{2} \left[ 2(1-m_{n}) I_{\rho}(s_{n}) -\tilde{H_r}(m_{n}) \right] + \Or(\sqrt{n}),
\end{equation}
which $\rho$ is the probability of observing a head state, and $I_{\rho}(s_{n})$ is the Bernoulli rate function \cite{touchette2009, Touchette2011}
\begin{equation}
I_{\rho}(x_2) = x_2 \ln (\frac{x_2}{\rho}) 
+  (1-x_2) \ln (\frac{1-x_2}{1-\rho}).
\end{equation}
The large deviation limit exists, and is equal to
\begin{equation}
\label{eq:ldp.lim.p_m_n_s_n}
\lim_{n \rightarrow \infty}  -\frac{1}{(n \ln n)} \ln P_n(m_{n}, s_{n}) = \frac{1-m_{n}}{2}.
\end{equation}

\section{Finding moments of the distribution $P_{2n}(n_p)$}
\label{appendix:moments}

First, let us find the first moment of $P_{2n}(n_p)$. Note that we use $2n$ instead of $n$ for the notation convenience, and later, write the final result for $n$.
The first moment of $P_{2n}(n_p)$ defines as
\begin{equation}
\langle n_p \rangle_{2n} = \sum_{i=0}^{n} i P_{2n}(n_p=i) = \frac{1}{c_{2n}(r)} \sum_{i=0}^{n} {n \choose 2i} (2i-1)!! \; i \; r^{n - i}.
\end{equation}
From (\ref{eq:f.2n.x})
\[
f_{2n}(x) = \sum_{i=0}^{n}  {2n \choose 2i} (2i-1)!! \; x^i \Rightarrow
\]
\begin{equation}
\left[x\frac{\rmd}{\rmd x}\right] f_{2n}(x) = \sum_{i=0}^{n}  {2n \choose 2i} (2i-1)!! \; i \; x^i,
\end{equation}
so
\begin{equation}
\langle n_p \rangle_{2n} = \frac{r^n}{c_{2n}(r)} \left. \left[x\frac{\rmd }{\rmd x}\right] f_{2n}(x) \right|_{x=\frac{1}{r}}.
\end{equation}
Furthermore, (\ref{eq:fn_x.by.derivative}) implies
\[
f_{2n}(x) := \left . \frac{\rmd^{2n} F(Y, x)}{\rmd Y^{2n}} \right|_{Y=0} \Rightarrow
\frac{\rmd f_{2n}(x)}{\rmd x} = \left . \frac{\rmd }{\rmd x}  \frac{\rmd^{2n} F(Y, x)}{\rmd Y^{2n}} \right|_{Y=0}
\]
\begin{equation}
= \left . \frac{\rmd^{2n} }{\rmd Y^{2n}} \frac{\rmd }{\rmd x} \exp(\frac{xY^2}{2}+Y) \right|_{Y=0} = \left . \frac{\rmd^{2n} }{\rmd Y^{2n}} \left(\frac{Y^2}{2} F(Y, x) \right) \right|_{Y=0},
\end{equation}
where in the step before the last one, we plugged in the generating function $F(Y, x)$ in (\ref{eq:gen.function.F.Y.x}). Taking the derivative of $F(Y, x)$ repetitively provides
\begin{equation}
\fl
\frac{\rmd^{2n} }{\rmd Y^{2n}} \left(\frac{Y^2}{2} F(Y, x) \right) = n(2n-1) F^{(2n-2)}(Y, x) + 2nYF^{(2n-1)}(Y, x) + \frac{Y^2}{2} F^{(2n)}(Y, x),
\end{equation}
which implies
\[
\frac{\rmd f_{2n}(x)}{\rmd x} = \left . \frac{\rmd^{2n} }{\rmd Y^{2n}} \left(\frac{Y^2}{2} F(Y, x) \right) \right|_{Y=0}
= n(2n-1) F^{(2n-2)}(0, x) 
\]
\begin{equation}
\label{eq:d_f_2n_dx}
= n(2n-1) f_{2n-2}(x).
\end{equation}
We can say, the derivative with respect to $x$ moves the polynomial $f_{2n}(x)$ to $f_{2n-2}(x)$ times $n(2n-1)$. Therefore,
\begin{equation}
\langle n_p \rangle_{2n} = \frac{r^n}{c_{2n}(r)} \left. x n(2n-1) f_{2n-2}(x) \right|_{x=\frac{1}{r}} 
= \frac{n(2n-1) }{c_{2n}(r)} r^{n-1} f_{2n-2}(\frac{1}{r}).
\end{equation}
But (\ref{eq:c_2n_both_cases}) defined $c_{2n-2}(r) = r^{n-1} f_{2n-2}(\frac{1}{r})$, thus
\begin{equation}
\label{eq:first.moment.even}
\langle n_p \rangle_{2n} = n(2n-1)\frac{c_{2n-2}(r)}{c_{2n}(r)}.
\end{equation}
This is the first moment for a system with even size $2n$. Similarly for 
an odd system size, we find 
\begin{equation}
\label{eq:first.moment.odd}
\langle n_p \rangle_{2n+1} = n(2n+1)\frac{c_{2n-1}(r)}{c_{2n+1}(r)}.
\end{equation}

Also $m_{n} = 2n_p/n$ is the ratio of the number of elements that are in a paired state to the system size $n$. Using the last results one obtains 
\begin{equation}
\fl
 \langle m_{2n} \rangle_r = \frac{2 \langle n_p \rangle_{2n}}{2n} = (2n-1)\frac{c_{2n-2}(r)}{c_{2n}(r)}, \qquad \langle m_{2n+1} \rangle_r = \frac{2 \langle n_p \rangle_{2n}}{2n+1} = 2n\frac{c_{2n-1}(r)}{c_{2n+1}(r)}.
\end{equation}
Combining both results asserts
\begin{equation}
\label{eq:first.moment.m_n}
\langle m_{n} \rangle_r = (n-1) \frac{c_{n-2}(r)}{c_{n}(r)}.
\end{equation}

To find the $k$-th moment, applying the operator $\left[x\frac{\rmd}{\rmd x}\right]$ $k$-times on $f_{2n}(x)$ results in
\begin{equation}
\left[x\frac{\rmd}{\rmd x}\right]^k f_{2n}(x) = \sum_{i=0}^{n}  {2n \choose 2i} (2i-1)!! \; i^k \; x^i,
\end{equation}
and the $k$-th moment, $\langle n_{p}^{k} \rangle_{2n}$, must be
\begin{equation}
\langle n_{p}^{k} \rangle_{2n} = \frac{r^n}{c_{2n}(r)} \left. \left[x\frac{\rmd}{\rmd x}\right]^k f_{2n}(x) \right|_{x=\frac{1}{r}}.
\end{equation}
In (\ref{eq:d_f_2n_dx}), we found the effect of single $\left[x\frac{\rmd}{\rmd x}\right]$ operator on $f_{2n}(x)$. Then, after careful applying the operator $\left[x\frac{\rmd}{\rmd x}\right]$ on $f_{2n}(x)$ for $k$ times and some bookkeepings, we obtain
\begin{equation}
\left[x\frac{\rmd}{\rmd x}\right]^k f_{2n}(x) = \sum_{i=1}^{k} \frac{n! (2n-1)!!}{(n-i)!(2n-2i-1)!!} a^{(k)}_{i} x^i f_{2n-2i}(x).
\end{equation}
The recursive equation for $a^{(k)}_{i}$ defines it as
\begin{equation}
a^{(k)}_{i} = a^{(k-1)}_{i-1} + ia^{(k-1)}_{i}, \qquad a^{(k)}_{1} = a^{(k)}_{k} = 1.
\end{equation}
Using (\ref{eq:c_2n_both_cases}) one writes $f_{2n-2i}(x)$ in terms of $c_{2n-2i}(r)$
\begin{equation}
f_{2n-2i}(\frac{1}{r}) = \frac{c_{2n-2i}(r)}{r^{n-i}},
\end{equation}
and therefore
\begin{equation}
\left. \left[x\frac{\rmd}{\rmd x}\right]^k f_{2n}(x) \right|_{x=\frac{1}{r}}  = \sum_{i=1}^{k}  a^{(k)}_{i} \frac{n! (2n-1)!!}{(n-i)!(2n-2i-1)!!} \frac{c_{2n-2i}(r)}{r^{n}},
\end{equation}
Finally, the $k$-th moment equation with respect to $P_{2n}(n_p)$ is 
\begin{equation}
\label{eq:k.moments}
\fl
\langle n_{p}^{k} \rangle_{2n} = \frac{r^n}{c_{2n}(r)} \left. \left[x\frac{\rmd}{\rmd x}\right]^k f_{2n}(x) \right|_{x=\frac{1}{r}}
= \sum_{i=1}^{k} a^{(k)}_{i} \frac{n! (2n-1)!!}{(n-i)!(2n-2i-1)!!} \frac{c_{2n-2i}(r)}{ c_{2n}(r)}.
\end{equation} 

\section{The $k$-th moment in terms of smaller size first moments}
\label{appendix:moments.sizes.relation}
Rewriting (\ref{eq:k.moments}), the $k$-th moment for a configuration with length $2n$ is
\begin{equation}
\langle n_{p}^{k}\rangle_{2n} := \sum_{i=1}^{k} b^{(k)}_{i} \frac{c_{2n-2i}(r)}{c_{2n}(r)}.
\end{equation}
where
\begin{equation}
b^{(k)}_{i}  = \frac{n! (2n-1)!!}{(n-i)!(2n-2i-1)!!} a^{(k)}_{i}.
\end{equation}
Observe 
\[
b^{(k)}_{i} \frac{c_{2n-2i}(r)}{ c_{2n}(r)} = \frac{n! (2n-1)!!}{(n-i)!(2n-2i-1)!!} a^{(k)}_{i} \frac{c_{2n-2i}(r)}{ c_{2n}(r)}
\]
\[
\fl
= \left[n(2n-1) \frac{c_{2n-2}(r)}{c_{2n}(r)} \right] \times\left[(n-1)(2n-3) \frac{c_{2n-4}(r)}{ c_{2n-2}(r)} \right]\times\left[(n-2)(2n-5) \frac{c_{2n-6}(r)}{ c_{2n-4}(r)} \right]\times \dots
\]
\[
\times \left[(n-i+1)(2n-2i+1) \frac{c_{2n-2i}(r)}{c_{2n-2i+2}(r)} \right] a^{(k)}_{i}
\]
\begin{equation}
= \langle n_{p} \rangle_{2n} \langle n_{p} \rangle_{(2n-2)} \langle n_{p} \rangle_{(2n-4)} \dots \langle n_{p} \rangle_{(2n-2i+2)} \times a^{(k)}_{i}.
\end{equation}
Therefore, the $k$-th moment for a system with size $2n$ relates to the first moments of systems with smaller sizes as
\begin{equation}
\label{eq:k.moment.to.smaller.size}
\langle n_{p}^{k} \rangle_{2n} = \sum_{i=1}^{k} a^{(k)}_{i} \langle n_{p} \rangle_{2n} \langle n_{p} \rangle_{(2n-2)} \langle n_{p} \rangle_{(2n-4)} \dots \langle n_{p} \rangle_{(2n-2i+2)}. 
\end{equation}

\section{Probability generating functions}
\label{appendix:generating.func}

For constant $n$, the number of pairs has an upper bound as $n_p \in \{0, 1 , \dots, n/2\}$. Hence, $P_{n}(n_p > n/2) = 0$. The probability generating function of $n_p$ with respect to $P_{n}(n_p)$ is
\[
G_{n}(s) = \sum_{n_p=0}^{\infty} P_{n}(n_p) s^{n_p}
= \frac{1}{c_{n}(r)} \sum_{n_p=0}^{\infty} {n \choose 2n_p} (2n_p-1)!! \; r^{\frac{n}{2} - n_p}  s^{n_p}
\]
\[
=  \frac{s^{\frac{n}{2}}}{c_{n}(r)} \sum_{n_p=0}^{\infty} {n \choose 2n_p} (2n_p-1)!! \; \left(\frac{r}{s}\right)^{\frac{n}{2} - n_p}  
\]
\begin{equation}
= s^{\frac{n}{2}} \frac{c_{n}(\frac{r}{s}) }{c_{n}(r)}.
\end{equation}
Notice that the normalization constant in the numerator evaluates at $(r/s)$.

Doing the same for $P_{n}(n_p, n_h)$ one finds
\[
G_{n}(s, u) = \sum_{n_p=0}^{\infty} \sum_{n_h=0}^{n-2n_p} P_{n}(n_p, n_h) s^{n_p} u^{n_h}
\]
\[
\fl
= \frac{1}{c_{n}(r)} \sum_{n_p=0}^{\infty} {n \choose 2n_p} (2n_p-1)!! \; r^{\frac{n}{2} - n_p}  s^{n_p}
\sum_{n_h=0}^{n-2n_p} {n-2n_p \choose n_h} \rho^{n_h} u^{n_h} (1-\rho)^{n-2n_p-n_h}
\]
\[
=  \frac{s^{\frac{n}{2}}}{c_{n}(r)} \sum_{n_p=0}^{\infty} {n \choose 2n_p} (2n_p-1)!! \; \left(\frac{r}{s}\right)^{\frac{n}{2} - n_p}  \left(\rho u + 1 - \rho \right)^{n-2n_p}
\]
\begin{equation}
= s^{\frac{n}{2}} \frac{c_{n}\left(\frac{r (\rho u + 1 - \rho )^{2}}{s}\right) }{c_{n}(r)}.
\end{equation}
The normalization constant in the numerator evaluates at $\left(r \left(\rho u + 1 - \rho \right)^{2}/s\right)$.

\section{The Marginal sums}
\label{appendix:marginal.sums}
Using the cardinality of $S_{i}$ in (\ref{eq:s_i_cardinality}), the marginal for $X_l = 1$ is
\[
P_{n}(X_l = 1)  = \sum_{i=0}^{n} P(\{S_{i}^{(1)}\}) 
= \frac{1}{c_{n}(r)} \sum_{i=0}^{n/2} {n-1 \choose 2i} (2i-1)!! \; r^{\frac{n}{2}-i}
\]
\[
= \frac{1}{c_{n}(r)} \sum_{i=0}^{n/2} {n \choose 2i} (2i-1)!! \; r^{\frac{n}{2}-i} - 
\frac{2}{n} \times \frac{1}{c_{n}(r)} \sum_{i=0}^{n/2} {n \choose 2i} (2i-1)!! \; i \; r^{\frac{n}{2}-i}
\]
\begin{equation}
= 1 - \frac{2 \langle n_p \rangle}{n} ,
\end{equation}
where we used the definition of the normalization constant and the expectation of $n_p$ in the last step. Similarly, using the cardinality of $S_{i}$ in (\ref{eq:s_i_k_cardinality}), the marginal for $X_l = 0$ is
\[
P_{n}(X_l = 0)  = \sum_{i=0}^{n/2} 
\sum_{\begin{array}{l}
	k=1 \\ 
	k \ne l
	\end{array}}^{n} P(\{S_{i,k}^{(2)}\})
\]
\[
= \frac{1}{c_{n}(r)} \sum_{i=0}^{n/2} 
\sum_{\begin{array}{l}
	k=1 \\ 
	k \ne l
	\end{array}}^{n} {n-2 \choose 2i-2} (2i-3)!! \; r^{\frac{n}{2}-i}
\]
\begin{equation}
= \frac{1}{c_{n}(r)} \sum_{i=0}^{n/2}  \frac{2i}{n} {n \choose 2i} (2i-1)!! \; r^{\frac{n}{2}-i} = \frac{2\langle n_p \rangle}{n} .
\end{equation}

\section{Asymptotic leading term of $\ln c_{n}(\epsilon)$}
\label{appendix:asymptotic.leading.term.espsilon}

It was mentioned that equation (\ref{eq:c.2n.asymptotic}) is a valid asymptotic leading term when $r$ is kept constant. Nonetheless, one special case happens when we assume $r$ is increasing with $n$ such that
\begin{equation}
\lim_{n \rightarrow \infty} \frac{r}{n} = \epsilon.
\end{equation}
The asymptotic term of $\ln c_{n}(\epsilon)$ needs considering this limit.
However, we start from the definition of $c_{n}(r)$ in (\ref{eq:c_n_r_def}) and write $c_{2n}(r)$ for $r = \epsilon (2n)$. Using an even system size, $2n$, makes notations uncluttered. Let us start by writing  $c_{2n}(r)$
\[
c_{2n}(r) = \sum_{i=0}^{n}  {2n \choose 2i} (2i-1)!! \; r^{n - i} 
\]
\begin{eqnarray}
= (2n)! \sum_{k=0}^{n} \frac{r^{k} 2^{k-n}}{ (n-k)! (2k)!},
\end{eqnarray}
where we used $k = n-2i$ in the last step. Replacing $r$ by $2\epsilon n$
\[
c_{2n}(\epsilon) = (2n)! \sum_{k=0}^{n} \frac{(2\epsilon n)^{k} 2^{k-n}}{ (n-k)! (2k)!}
\]
\begin{equation}
\label{eq:appendix.c_2n_r}
= \frac{(2n)!}{2^n} \sum_{k=0}^{n} \frac{(4\epsilon n)^{k} }{ (n-k)! (2k)!}.
\end{equation}
Without rearranging the order of $k$, in the limit $2n \rightarrow \infty$, $k_{max}$ or where the summand is maximum approaches to $n$. Yet, in the above form, $k_{max}$ is $\epsilon$ dependent. 
Denoting the summand by
\begin{equation}
t_{2n}(k) = \frac{(4\epsilon n)^{k} }{ (n-k)! (2k)!},
\end{equation}
and using the Sterling's approximation $ \ln (n!) \sim n \ln n - n$, we find the $k_{max}(\epsilon)$ by taking the derivative of logarithm of $t_{2n}(k)$. Since a logarithm is a strictly increasing function, the maximum of $t_{2n}(k)$ coincides with $\ln t_{2n}(k)$
\[
\ln t_{2n}(k) = k \ln (4\epsilon n) - 2k \ln (2k) + 2k - (n-k) \ln(n-k) + (n-k) \Rightarrow
\]
\[
\frac{d \ln t_{2n}(k)}{dk} = \ln (4\epsilon n) - 2 - 2 \ln (2k) + 2 + 1 + \ln(n-k) -1
\]
\[
= \ln (4\epsilon n) + \ln\frac{n-k}{4k^2} = 0 \Rightarrow
\]
\begin{equation}
k^2 + \epsilon n k - \epsilon n^2 = 0.
\end{equation}
The solutions of the above quadratic equation are
\begin{equation}
k = -\frac{\epsilon n}{2} \pm \frac{\epsilon n}{2} \sqrt{1 + \frac{4}{\epsilon}}.
\end{equation}
However, $k$ is non-negative, and therefore, the maximum of $t_{2n}(k)$ is at
\begin{equation}
k_{max} = \frac{\epsilon n}{2} \left[\sqrt{1 + \frac{4}{\epsilon}} - 1 \right].
\end{equation}
To not clutter the notation, we denote the bracket as
\begin{equation}
\label{eq:c_2n_epsilon_f}
f(\epsilon) = \sqrt{1 + \frac{4}{\epsilon}} - 1,
\end{equation}
and write
\begin{equation}
k_{max} = \frac{\epsilon n f(\epsilon)}{2}.
\end{equation}

The sum in the definition of $c_{2n}(\epsilon)$ is concentrated at its maximum. We find the asymptotic leading term of $t_{2n}(k)$ as follows: first simplifying it by using the Sterling's approximation, $ n! = \sqrt{2 \pi n} \left(n/\rme\right)^n$, and next evaluating it at $k_{max}$
\[
t_{2n}(k) = \frac{(4\epsilon n)^{k} }{ (n-k)! (2k)!} = \frac{ (4\epsilon n)^{k} \rme^{n+k}}{ 2 \pi \sqrt{2k(n-k)} (2k)^{2k} (n-k)^{n-k}}
\]
\begin{equation}
=  \frac{1}{2 \sqrt{2} \pi } \times \frac{1}{\sqrt{k(n-k)}} \times \frac{1}{  (1-\frac{k}{n})^{n-k}} \times (\frac{\rme \epsilon n^2}{k^2})^{k} \left(\frac{\rme}{n}\right)^{n}.
\end{equation}

The first ratio $1/\sqrt{k(n-k)}$ approximates as
\begin{equation}
\frac{1}{\sqrt{k_{max}(n-k_{max})}} \sim \frac{1}{n \sqrt{\frac{\epsilon f(\epsilon)}{2}  (1- \frac{\epsilon f(\epsilon)}{2})}},
\end{equation}
considering the fact that the bulk of the distribution is concentrated around the $k_{max}$. Let us define
\begin{equation}
\label{eq:c_2n_epsilon_g}
g(\epsilon) = (1- \frac{\epsilon f(\epsilon)}{2}).
\end{equation}
Then
\begin{equation}
\frac{1}{\sqrt{k(n-k)}} \sim \frac{1}{n \sqrt{\frac{\epsilon}{2} f(\epsilon) g(\epsilon)}},
\end{equation}
and
\begin{equation}
(1-\frac{k}{n})^{n-k} \sim   (1-\frac{\epsilon f(\epsilon)}{2})^{n(1-\frac{\epsilon f(\epsilon)}{2})} = g(\epsilon)^{n g(\epsilon)}.
\end{equation}
Therefore, the summand turns to
\begin{equation}
t_{2n}(k) \sim  \frac{1}{2 \sqrt{2} \pi } \times \frac{1}{n \sqrt{\frac{\epsilon}{2} f(\epsilon) g(\epsilon)}} \times  g(\epsilon)^{-n g(\epsilon)} \times (\frac{\rme \epsilon n^2}{k^2})^{k} \left(\frac{\rme}{n}\right)^{n},
\end{equation}
and $c_{2n}(\epsilon)$ becomes
\begin{equation}
c_{2n}(\epsilon) \sim  \frac{(2n)!}{2^{n+1} \pi} \frac{g(\epsilon)^{-n g(\epsilon)} \left(\frac{\rme}{n}\right)^{n}}{n \sqrt{\epsilon f(\epsilon) g(\epsilon)}} \sum_{k=1}^{n} (\frac{\rme \epsilon n^2}{k^2})^{k}.
\end{equation}
We have to emphasis that we started the sum from $k=1$ instead of $k=0$. To justify it, observe that using (\ref{eq:appendix.c_2n_r}) we derive the ration of two consecutive summands for $k=0$ and $k=1$
\begin{equation}
\frac{t_{2n}(0)}{t_{2n}(1)} = \frac{1}{4 \epsilon n} \times \frac{(n-1)!}{n!} \times \frac{0!}{2!}
= \frac{1}{8 \epsilon n^2}.
\end{equation}
Asymptotically it means $t_{2n}(0) \ll t_{2n}(1)$ for any $\epsilon > 0$ as $n \rightarrow \infty$. And it justifies the exclusion of $k=0$. 
If we approximate $(2n)!$, it results in
\[
c_{2n}(\epsilon) \sim  \frac{\sqrt{4 \pi n} (\frac{2n}{\rme})^{2n} \left(\frac{\rme}{n}\right)^{n}}{2^{n+1} \pi} \frac{g(\epsilon)^{-n g(\epsilon)} }{n \sqrt{ \epsilon f(\epsilon) g(\epsilon)}} \sum_{k=1}^{n} (\frac{\rme \epsilon n^2}{k^2})^{k}
\]
\begin{equation}
= \frac{(\frac{2n}{\rme})^{n} g(\epsilon)^{-n g(\epsilon)} }{ \sqrt{n \pi \epsilon f(\epsilon) g(\epsilon)}} \sum_{k=1}^{n} (\frac{\rme \epsilon n^2}{k^2})^{k}.
\end{equation}
Finally, we need to estimate the asymptotic leading term of the sum $\sum_{k=0}^{n} (\sqrt{\rme \epsilon} n/k)^{2k}$. Similar to the argument that reported in \cite{jensen2018}, the sum can be estimated as an integral. Still, the free parameter $\epsilon$ introduces two different behaviours that needs a subtle consideration.
Let us call the summand as
\begin{equation}
\psi(k) = (\frac{\sqrt{\rme \epsilon} n}{k})^{2k}.
\end{equation}
Similar to what we have already done to find $k_{max}$, we treat $\psi(k)$ as a continuous function. And to find its maximum with respect to $\epsilon$, we take the derivative
\[
\frac{d \ln \psi(k)}{d k} = 2 \log (\frac{\sqrt{\rme \epsilon} n}{k}) - 2 = 0 \Rightarrow
\]
\begin{equation}
\frac{k_{max}}{n} = \sqrt{\frac{\epsilon}{\rme}}.
\end{equation}
However, $k_{max} \le n$, and for $\epsilon > e$ the above result implies 
\begin{equation}
k_{max} = n.
\end{equation}
In other words, the maximum of the sum is concentrated at $k_{max} = n$. Thus the bound $\epsilon > \rme$ must be considered in finding the asymptotic leading term.
Continuing the estimation by an integral, we divide the range of $\epsilon$ to two regions
\begin{itemize}
	\item $\epsilon > \rme$: In this case the sum can be estimated by its largest term 
	\begin{equation}
	\sum_{k=1}^{n} (\frac{\rme \epsilon n^2}{k^2})^{k} \sim (\frac{\rme \epsilon n^2}{n^2})^{n}  = \left(\rme \epsilon \right)^{n}.
	\end{equation}

	\item $\epsilon \le \rme$: We have 
	\[
	\sum_{k=1}^{n} (\frac{\rme \epsilon n^2}{k^2})^{k} \sim \int_{1}^{n} \left(\frac{ \sqrt{\rme \epsilon} n}{x}\right)^{2x} dx
	\]
	\[
	= \sqrt{\rme \epsilon} \; n \int_{\frac{1}{\sqrt{\rme \epsilon} n}}^{\frac{1}{\sqrt{\rme \epsilon}}} x^{-2\sqrt{\rme \epsilon}\; nx} dx 
\qquad (x \rightarrow \sqrt{\rme \epsilon} \;  n x)
	\]
	\begin{equation}
	= \sqrt{\rme \epsilon} \;  n \int_{\frac{1}{\sqrt{\rme \epsilon} n}}^{\frac{1}{\sqrt{\rme \epsilon}}} \rme^{-n\left[2\sqrt{\rme \epsilon}\;  x \ln x \right]} dx.
	\end{equation}
	We are going to use the steepest decent approximation here. Define
	\begin{equation}
	h(x) = 2\sqrt{\rme \epsilon} \; x \ln x,
	\end{equation}
	and see that $h'(x^{*}) = 0$ gives
	\begin{equation}
	x^{*} = \rme^{-1}.
	\end{equation}
	So, the Taylor expansion of $h(x)$ around $x^{*}$, up and including the quadratic term, is
	\[
	h(x) = h(x^{*}) + \frac{h''(x^{*})}{2}(x-x^{*})^2 + O(x^3)  
	\]
	\begin{equation}
	= -2\sqrt{\frac{\epsilon}{\rme}} + \sqrt{\rme^{3} \epsilon } (x-\rme^{-1})^2 + O(x^3).
	\end{equation}
	Then, defining $a^2 \equiv \sqrt{\rme^{3} \epsilon }$, the integral estimate gives
	\[
	\sum_{k=1}^{n} (\frac{\rme \epsilon n^2}{k^2})^{k} \sim 
	\sqrt{\rme \epsilon} \; n \int_{\frac{1}{\sqrt{\rme \epsilon} n}}^{\frac{1}{\sqrt{\rme \epsilon}}} \rme^{-n\left[-2\sqrt{\frac{\epsilon}{\rme}} + a^2 (x-\rme^{-1})^2 \right]} dx
	\]
	\[
	= \frac{\sqrt{\rme \epsilon}}{a \sqrt{n}} \; n \rme^{2n\sqrt{\frac{\epsilon}{\rme}}} \int_{a \sqrt{n} (\frac{1}{\sqrt{\rme \epsilon} n}-\rme^{-1})}^{a\sqrt{\frac{n}{\rme}}(\frac{1}{\sqrt{\epsilon}}-\frac{1}{\sqrt{\rme}})} \rme^{ - t^2 } dt
	\qquad (t = a \sqrt{n} (x-\rme^{-1}))
	\]	
	\begin{equation}
	= \left(\frac{\epsilon}{\rme}\right)^{\frac{1}{4}} \sqrt{n} \; \rme^{2n\sqrt{\frac{\epsilon}{\rme}}} \int_{- \left[ \left(\frac{\epsilon}{\rme}\right)^{\frac{1}{4}} - \left(\frac{\rme}{\epsilon}\right)^{\frac{1}{4}} \frac{1}{n} \right] \sqrt{n}}^{   \left[ \left(\frac{\rme}{\epsilon}\right)^{\frac{1}{4}} - \left(\frac{\epsilon}{\rme}\right)^{\frac{1}{4}} \right] \sqrt{n} } \rme^{ - t^2 } dt.
	\end{equation}

	For $n \rightarrow \infty$ and considering the condition $\epsilon \le \rme$, the limits of the integral become
	\begin{equation}
	\lim_{n \rightarrow \infty} - \left[ \left(\frac{\epsilon}{\rme}\right)^{\frac{1}{4}} - \left(\frac{\rme}{\epsilon}\right)^{\frac{1}{4}} \frac{1}{n} \right] \sqrt{n} \rightarrow -\infty,
	\end{equation}
	and
	\begin{equation}
	\lim_{n \rightarrow \infty}    \left[ \left(\frac{\rme}{\epsilon}\right)^{\frac{1}{4}} - \left(\frac{\epsilon}{\rme}\right)^{\frac{1}{4}} \right] \sqrt{n} \rightarrow \infty.
	\end{equation}
	Then
	\begin{equation}
	\fl
	\sum_{k=1}^{n} (\frac{\rme \epsilon n^2}{k^2})^{k} \sim \left(\frac{\epsilon}{\rme}\right)^{\frac{1}{4}} \sqrt{n} \; \rme^{2n\sqrt{\frac{\epsilon}{\rme}}} \int_{-\infty}^{\infty} \rme^{ - t^2 } dt
	= \left(\frac{\epsilon}{\rme}\right)^{\frac{1}{4}} \sqrt{\pi n} \; \rme^{2n\sqrt{\frac{\epsilon}{\rme}}}.
	\end{equation}

\end{itemize}

So, for both regions we must have
\begin{equation}
\sum_{k=0}^{n} (\frac{\rme \epsilon n^2}{k^2})^{k} \sim
=  \left\lbrace  \begin{array}{ll}
\left(\frac{\epsilon}{\rme}\right)^{\frac{1}{4}} \sqrt{\pi n} \; \rme^{2n\sqrt{\frac{\epsilon}{\rme}}} & \epsilon \le \rme\\
(\epsilon \rme)^n & \epsilon > \rme
\end{array}\right. .
\end{equation}

Finally, the $c_{2n}(\epsilon)$ asymptotic leading term is
\[
c_{2n}(\epsilon) \sim \frac{(\frac{2n}{\rme})^{n} g(\epsilon)^{-n g(\epsilon)} }{ \sqrt{n \pi \epsilon f(\epsilon) g(\epsilon)}} \times 
\left\lbrace  \begin{array}{ll}
\left(\frac{\epsilon}{\rme}\right)^{\frac{1}{4}} \sqrt{\pi n} \; \rme^{2n\sqrt{\frac{\epsilon}{\rme}}} & \epsilon \le \rme\\
(\epsilon \rme)^n & \epsilon > \rme
\end{array}\right.
\]
\begin{equation}
\label{eq:c_2n_epsilon}
= 
\left\lbrace  \begin{array}{ll}
\frac{g(\epsilon)^{-n g(\epsilon)} \rme^{2n\sqrt{\frac{\epsilon}{\rme}} }  (\frac{2n}{\rme})^{n}}{ \sqrt{ \sqrt{\rme\epsilon}  f(\epsilon) g(\epsilon)}} & \epsilon \le \rme\\
\frac{g(\epsilon)^{-n g(\epsilon)} (2 \epsilon n)^{n}}{\sqrt{n \pi \epsilon f(\epsilon) g(\epsilon)}} & \epsilon > \rme
\end{array}\right..
\end{equation}

\section*{References}
\bibliography{iopart-num}

\end{document}